\documentstyle[12pt,rotating,epsf]{article}
\newcommand{\Frac}[2]{\frac{\displaystyle #1}{\displaystyle #2}}
\newcommand{\kpiggp}{$K^+ \rightarrow \pi^+ \gamma \gamma $ }
\newcommand{\kpiggtot}{$K \rightarrow \pi \gamma \gamma $ }

\newcommand{\kpiggn}{$K_L \rightarrow \pi^0 \gamma \gamma $ }
\newcommand{\kpiggne}{K_L \rightarrow \pi^0 \gamma \gamma  }
\newcommand{\kgll}{$K_L \rightarrow  \gamma \ell^+ \ell^- $ }
\newcommand{\kgee}{$K_L \rightarrow  \gamma e^+ e^- $ }
\newcommand{\kgmm}{$K_L \rightarrow  \gamma \mu^+ \mu^- $ }
\newcommand{\kggs}{$K_L \rightarrow  \gamma \gamma^* $ }
\newcommand{\opc}{${\cal O}(p^4)$ }
\newcommand{\ops}{${\cal O}(p^6)$ }

\def\mapright#1{\smash{
     \mathop{\longrightarrow}\limits^{#1}}}
\begin{document}
\pagestyle{empty}
\begin{titlepage}
\begin{center}
\vspace*{-1.1cm}
\hfill  INFNNA-IV-96/21\\
\hfill  DSFNA-IV-96/21\\
\hfill September, 1996
\vspace*{1.4cm} \\
{\LARGE \bf $^*$ Vector meson exchange contributions to  \\ 
\vspace*{0.5cm} $K \rightarrow \pi \gamma \gamma$ and $K_L \rightarrow
\gamma \ell^+ \ell^-$ \\ }
\vspace*{1.6cm}
{\large \bf Giancarlo D'Ambrosio$^{\dagger}$} $ \; \; $ and $ \; \; $
{\large \bf  Jorge Portol\'es$^{\ddagger}$ }
\vspace*{0.4cm} \\
Istituto Nazionale di Fisica Nucleare, Sezione di Napoli \\
Dipartamento di Scienze Fisiche, Universit\`a di Napoli \\
I-80125 Napoli, Italy \\ 
\vspace*{1.8cm} 

\begin{abstract}
We have studied in the framework of Chiral Perturbation Theory ($\chi$PT)
the ${\cal O}(p^6)$ Vector Meson contributions to $K \rightarrow
\pi \gamma \gamma$ and $K_L \rightarrow \gamma \gamma^*$. We construct
the most general $\chi$PT lagrangian for the weak Vector--Pseudoscalar--Photon
($VP\gamma$) vertex for $K \rightarrow \pi \gamma \gamma$ and 
$K_L \rightarrow \gamma \gamma^*$
and consequently we get the full structure of the
${\cal O}(p^6)$ local contributions generated by Vector Meson exchange.
Then we compute new factorizable contributions to the weak $VP\gamma$ 
vertex generated by the odd--intrinsic parity violating lagrangian with 
no additional couplings. We find in fact
a very nice agreement with the phenomenology of $K \rightarrow \pi 
\gamma \gamma$ and $K_L \rightarrow \gamma \gamma^*$, more predictive
power and a deeper understanding of the ${\cal O}(p^6)$ local operators. 
A novel 
interpretation of the $K_L \rightarrow \gamma \gamma^*$ data is given. 
Also a comparison with the existing models is presented.
\end{abstract}
\end{center}
\vspace*{0.4cm}
PACS~: 12.15.-y,12.39.Fe,12.40.Vv,13.25.Es \\
Keywords~: Radiative non--leptonic kaon decays, Non--leptonic weak
hamiltonian, Chiral Perturbation Theory, Vector meson dominance.
\vspace*{0.4cm}\\
$^{\dagger}$ E-mail~: dambrosio@axpna1.na.infn.it \\
$^{\ddagger}$ E-mail~: portoles@axpna1.na.infn.it \\
\vfill
\noindent * Work supported in part by HCM, EEC--Contract No. 
CHRX--CT920026 (EURODA$\Phi$NE).
\end{titlepage}
\newpage
\pagestyle{plain}
\pagenumbering{arabic}

\section{Introduction}
\hspace*{0.5cm}
Our knowledge of the phenomenology of radiative non--leptonic kaon decays
both from theoretical and experimental points of view has improved
drastically in the last ten years. The main tool that has ushered this
progress has been the application of Chiral Perturbation Theory 
($\chi$PT) \cite{WE79,GL85,MG84} techniques to weak processes. $\chi$PT is 
a quantum 
effective theory that satisfies the basic chiral symmetry of QCD and
supports a perturbative Feynman--Dyson expansion in masses and external
momenta. Correspondingly at any new order in the expansion new chiral 
invariant operators,
with coupling constants not determined by the symmetry, appear. At 
\opc vector meson exchange (when present) has been shown to be 
effective in predicting those couplings in the strong sector 
\cite{EG89,DR89}. This is not an easy task, however, in the weak 
lagrangian since the weak couplings of vector mesons are much less known.
Thus various models implementing weak interactions
at the hadronic level have been proposed, yet it is very likely that 
mechanisms and couplings working for a subset of processes might not work
in general for other processes. Nevertheless the information provided
by the models can be useful in giving a general picture of the 
hadronization process of the interaction.
\par 
$\chi$PT has been successfully used to study
 radiative non--leptonic kaon decays
(see \cite{DE95,EDU95,DAGI96} and references therein). 
Here we are concerned in the study of the ${\cal O}(p^6)$ vector meson 
exchange contributions to \kgll, \kpiggn and \kpiggp decays.
\par
The \kggs form factor (together with $K_L \rightarrow \gamma^* \gamma^*$)
is an important ingredient in order to evaluate properly the dispersive
contribution to the real part of the amplitude for the decay
$K_L \rightarrow \gamma^* \gamma^* \rightarrow \mu^+ \mu^-$. Since the
real part receives also short distance contributions proportional
to the CKM matrix element $V_{td}$ \cite{NEB77} and the absorptive 
amplitude is found to saturate the experimental result, a strong 
cancellation in the real part between short and long distance 
contributions is expected. Analysis of the relevant form factor 
in $K_L \rightarrow \gamma \gamma^*$ \cite{BMS83} shows a 
relatively large weak \ops vector meson contribution and, therefore,
it is of concern for our study here.
\par
The importance of \kpiggn goes further the interest of the process in 
itself because its r\^ole as a CP--conserving two--photon discontinuity
amplitude to $K_L \rightarrow \pi^{0} e^+ e^-$ in possible 
competition with the CP--violating contributions \cite{DH87,SE88,FR89}. 
The leading \opc $\chi$PT amplitude
to \kpiggn, which does not receive any local contribution, though 
predicting well the diphoton invariant mass spectrum, underestimates
by a factor 3 the rate. No complete \ops computation has been made;
however, there are large unitarity corrections \cite{CD93,CE93} and one
could fix the weak coupling carrying the information of the vector 
meson exchange to reproduce rate and spectrum \cite{CE93}. This weak
vector coupling 
also turns out to be large and it is not well explained by the theoretical
models that have been used until now.
\par
The interest on \kpiggp relies also on its actuality. BNL-787 has got,
for the first time \cite{TAK96}, 31 preliminary events on this channel 
and KLOE at 
DA$\Phi$NE will do in the near future. This good experimental perspectives
will give information about the \opc weak counterterms in the chiral 
lagrangian since now the \ops unitarity corrections have already been evaluated
\cite{DP96}. At this chiral order is then necessary to have a control
of the vector exchange contributions.
\par
The interplay between experimental results and phenomenology indicates that
in both these decays ($K \rightarrow \pi \gamma \gamma$
and $K_L \rightarrow \gamma \gamma^*$) there is an important \ops vector
meson exchange contribution, which prediction, as commented above, relies
on theoretical models. We will review them in the
paper. Let us collect here their main achievements :
\begin{itemize}
\item[i)] Factorization Model (FM) \cite{PI91}.- Motivated by $1/N_c$ 
models can give a consistent picture of both \kpiggn and \kggs, according
to experiment, if the effective coupling (a free parameter in this model)
is carefully chosen as we will show.
\item[ii)] Bergstr\"om--Mass\'o--Singer model (BMS) \cite{BMS83} .- For 
$K_L \rightarrow \gamma \gamma^*$ assumes the sequence $K_L \rightarrow
K^* \gamma$; $K^* \rightarrow \rho,\omega,\phi$; $\rho,\omega,\phi \rightarrow
\gamma^*$ where the weak $K^* \rightarrow \rho,\omega,\phi$ transition 
is computed using vacuum insertion. In the BMS model it is also assumed
that the weak vector--vector couplings 
are given by the perturbative Wilson coefficients (i.e. no 
$\Delta I = 1/2$ enhancement). This model gives good results for 
$K_L \rightarrow \gamma \gamma^*$ but we will show 
that it produces a too small result
for the vector meson contribution to $K_L \rightarrow \pi^{0}
\gamma \gamma$.
\item[iii)] Weak Deformation Model (WDM) \cite{EP90,ECG90} .- It might 
reproduce well the 
$K_L \rightarrow \gamma \gamma^*$ slope but gives a too small vector
meson  contribution
to $K_L \rightarrow \pi^{0} \gamma \gamma$.
\end{itemize}
Actually the experimental determination of the slope in \kggs is only
based in the BMS model. We analyse critically this experimental result
and propose an alternative analysis less model dependent.
\par
We realize also that the common problematic point between \kpiggtot and 
\kggs is the model dependence in the construction of the weak 
$VP\gamma$ vertex. Thus we have constructed the most general $\chi$PT
lagrangian for the weak $VP\gamma$ vertex for the processes under
consideration. As we will show this gives us a quantitative relation
between the two relevant parameters in \kpiggtot and \kggs, thus correlating 
both processes.
\par
Furthermore we propose a Factorization Model 
in the Vector couplings (FMV) that has as main ingredients :
\begin{itemize}
\item[a)] The application of the FM to the construction of the weak $VP\gamma$
vertex instead of the effective $K\pi\gamma\gamma$ and $K\gamma\gamma^*$
vertices as in the usual approach.
\item[b)] We consider that the effective coupling of the FMV model 
(the only free parameter) is given by the Wilson coefficient in the
non--leptonic hamiltonian.
\end{itemize}
We confront this model with the phenomenology and with the 
aforementioned models, particularly with the FM. Also we do a comparison
with Ko's reference \cite{KO91} where similar questions have been
addressed.
\par
The structure of the paper is as follows. In Section 2 we discuss briefly
the treatment of non--leptonic weak interactions at low energy.
In Section 3 we specify the 
kinematics and notation for our processes of interest. In Section 4 we 
construct the most general chiral
structure of the octet piece for the weak $VP\gamma$ vertex, we also
collect the lagrangian densities and the parameterizations we will 
need in our study and we specify
our conventions; finally we discuss
the predictions of the FM, BMS and WDM models for the observables in 
\kpiggtot and \kggs. In Section 5 we propose our model in order to do a
more complete evaluation of the factorizable contributions to the 
amplitudes. Results and conclusions of our work will be discussed
in Sections 6 and 7 respectively.

\section{Non--leptonic weak interactions at low energies}
\hspace*{0.5cm}
We will review in this Section the procedures of $\chi$PT and their 
implementation in the study of non--leptonic weak processes. At the same
time we will introduce our notations and the tools we will need in the
development of our study.

\subsection{Chiral Perturbation Theory}
\hspace*{0.5cm}
$\chi$PT \cite{WE79,GL85} is an effective quantum field theory for the study
of low energy strong interacting processes that relies in the exact 
symmetry of massless QCD. 
The basic 
assumption of $\chi$PT is that, at low energies ($E \leq \, 1 \, GeV$), 
the chiral symmetry group $G \equiv SU(3)_L \otimes SU(3)_R$
is spontaneously broken to the vector subgroup
$SU(3)_V$. 
\par
Following the basic references by Callan, Coleman, Wess and Zumino 
\cite{CCW69} we introduce $u(\varphi)$ 
as an element of the coset space $G/H$, with $H=SU(3)_V$, that transforms
under the chiral group as
\begin{equation}
u(\varphi) \, \mapright{G} \, g_R \, u(\varphi) \, h(g,\varphi)^{\dagger}
\, = \, h(g,\varphi) \, u(\varphi) \, g_L^{\dagger} \; \; \; , 
\label{eq:urht}
\end{equation}
where $h(g,\varphi) \, \in \, H$ is a compensator field.
A convenient parameterization of $u(\varphi)$ is given by
\begin{equation}
u (\varphi) \, = \, \exp \left( \, \Frac{i}{2 \, F} \sum_{j=1}^8 
\, \lambda_j \varphi_j \, \right) \; \; \; \; , 
\label{eq:uphi}
\end{equation}
where $\lambda_i$ are the $SU(3)$ Gell--Mann matrices \footnote{Normalized
to $Tr(\lambda_i \lambda_j) = 2 \delta_{ij}$.} and $F \, \sim \, F_{\pi} 
\, \simeq \, 93 \, MeV$ is the decay constant of the pion.
\par
The extension of the global chiral symmetry to a local one and the inclusion
of external fields are convenient tools in order to work out systematically
the Green functions of interest and the construction of chiral operators
in the presence of symmetry breaking terms. A covariant derivative on 
the $U(\varphi) \, = \, u(\varphi) u(\varphi)$ field is then defined as
\begin{equation}
D_{\mu} U \, = \, \partial_{\mu} U \, 
- \, i r_{\mu} U \, + \, i U \ell_{\mu} \; \; \; ,
\label{eq:ducov}
\end{equation}
where $\ell_{\mu} \, = \, v_{\mu} \, - \, a_{\mu}$ and $r_{\mu} \, = \, 
v_{\mu} \, + \, a_{\mu}$, in terms of the external vector and axial fields,
are the left and right external fields, respectively. If only the 
electromagnetic field is considered then
$\ell_{\mu} \, = \, r_{\mu} \, = \, - \, e \, Q \, A_{\mu}$ where
$Q \equiv diag(2/3 \, , \, -1/3 \, , \, -1/3)$ is the electric charge
matrix of the $u,d$ and $s$ quarks
\footnote{This corresponds to $D_{\mu} \phi^{\pm} \, = \, (\partial_{\mu} 
\pm i e A_{\mu}) \phi^{\pm}$.} . The explicit symmetry breaking of the
chiral symmetry due to the masses of the octet of pseudoscalars is 
included through the external scalar field $ s \, = \, {\cal M} \, + ...$
In this way an effective lagrangian can be 
constructed as an expansion in the external momenta (derivatives of the
Goldstone fields) and masses \cite{WE79,GL85,MG84}.
The leading 
${\cal O} (p^2)$ strong lagrangian is
\begin{equation}
{\cal L}_2 \, = \, \Frac{F^2}{4} \, \langle \, u_{\mu} \, 
u^{\mu} \, + \, \chi_{+} \,  \rangle \; \; \; , 
\label{eq:str2}
\end{equation}
where $\langle \, A \, \rangle  \, \equiv \, Tr(A)$ in the flavour space,
and
\begin{eqnarray}
u_{\mu} \, \equiv \, i \, u^{\dagger} \, D_{\mu} \, U \, u^{\dagger} 
\; \; \;  & , & \; \;  u_{\mu} \, \mapright{G} \, h(g,\varphi) \, 
u_{\mu} \, h(g,\varphi)^{\dagger} \; \; , \nonumber \\
\chi_{+} \, \equiv  \, u^{\dagger} \, \chi \, u^{\dagger} \, + \, 
u \, \chi^{\dagger} \, u \; \; \; & , & \; \; 
\chi \, \equiv \, 2 \, B_{\circ} \, ( s \, + \, i \, p) \, \simeq \, 
2 \, B_{\circ} \, {\cal M} \, + ... \; \; , 
\label{eq:extrasym} \\
{\cal M} \, \equiv \, diag(m_u \, , \, m_d \, , \, m_s) \; \; \; & , & \;
\; B_{\circ} \, \equiv \, - \, \Frac{1}{F^2} \, \langle 0 | \,
\overline{u} u \, |0 \rangle \; \; \; . \nonumber
\end{eqnarray}
The even--intrinsic parity lagrangian 
at \opc was developed in \cite{GL85} and introduces 12 new
coupling constants.
The \opc odd--intrinsic parity lagrangian arises as a solution to the
Ward condition imposed by the chiral anomaly \cite{WZW71}. The chiral
anomalous functional $Z_{an} [U,\ell, r]$ as given by the 
Wess--Zumino--Witten action (WZW) is
\begin{eqnarray}
Z_{an}[U,\ell,r]_{WZW} \, & = & - \, \Frac{i N_c}{240 \pi^2} \, 
\int_{M^5} \, d^5x \epsilon^{ijklm}\langle \Sigma_i^L \Sigma_j^L
\Sigma_k^L \Sigma_l^L \Sigma_m^L \rangle \nonumber \\
& &  
\label{eq:zulr} \\
& & - \, \Frac{i N_c}{48 \pi^2} \, 
\int \, d^4x \varepsilon_{\mu \nu \alpha \beta} \, 
( W(U,\ell,r)^{\mu \nu \alpha \beta} \, - \, 
W(I,\ell,r)^{\mu \nu \alpha \beta} \, ) \; \; \; , \nonumber
\end{eqnarray}
\begin{eqnarray}
W(U,\ell,r)_{\mu \nu \alpha \beta} \, & = & \langle
U \ell_{\mu} \ell_{\nu} \ell_{\alpha} U^{\dagger} r_{\beta} \, + \, 
\Frac{1}{4}U \ell_{\mu} U^{\dagger} r_{\nu} U \ell_{\alpha} 
U^{\dagger} r_{\beta} \, + \, i U \partial_{\mu} \ell_{\nu} \ell_{\alpha}
U^{\dagger} r_{\beta} \nonumber \\
& & \; + \, i \partial_{\mu} r_{\nu} U \ell_{\alpha} U^{\dagger} r_{\beta} \,
- \, i \Sigma_{\mu}^L \ell_{\nu} U^{\dagger} r_{\alpha} U \ell_{\beta} \, 
+ \, \Sigma_{\mu}^L U^{\dagger} \partial_{\nu} r_{\alpha} U \ell_{\beta} 
\nonumber \\
& & \; - \, \Sigma_{\mu}^L \Sigma_{\nu}^L U^{\dagger} r_{\alpha} U 
\ell_{\beta} \, + \, \Sigma_{\mu}^L \ell_{\nu} \partial_{\alpha} \ell_{\beta}
\, + \, \Sigma_{\mu}^L \partial_{\nu} \ell_{\alpha} \ell_{\beta} 
\nonumber \\
& & \; - \, i \Sigma_{\mu}^L \ell_{\nu} \ell_{\alpha} \ell_{\beta} \, 
+ \, \Frac{1}{2} \Sigma_{\mu}^L \ell_{\nu} \Sigma_{\alpha}^L \ell_{\beta}
\, - \, i \Sigma_{\mu}^L \Sigma_{\nu}^L \Sigma_{\alpha}^L \ell_{\beta} 
\, \rangle
\nonumber \\
& & \; - \; ( \, L \, \leftrightarrow \, R \, ) \; \; \; , 
\label{eq:wulr} 
\end{eqnarray}
with $N_c = 3$, $\Sigma_{\mu}^L \, = \, U^{\dagger} \partial_{\mu} U$,
$\Sigma_{\mu}^R \, = \, U \partial_{\mu} U^{\dagger}$
and $(L \leftrightarrow R)$ stands for the interchange
$U \leftrightarrow U^{\dagger}$,  
$\ell_{\mu} \leftrightarrow r_{\mu}$, 
$\Sigma_{\mu}^L \leftrightarrow \Sigma_{\mu}^R$.
We see that the WZW action does not introduce any unknown coupling.
\par
The inclusion of other quantum fields than the pseudoscalar Goldstone bosons
in the chiral lagrangian was also considered in \cite{CCW69}. We are 
interested in the introduction of vector mesons coupled to the $U(\varphi)$
and to the external fields. Let us then introduce the nonet of vector 
fields
\begin{equation}
V_{\mu} \, = \, \Frac{1}{\sqrt{2}} \, \sum_{i=1}^{8} \, \lambda_i \, 
V_{\mu}^i \, + \, \Frac{1}{\sqrt{3}} \, V_{\mu}^0 \; \; \; , 
\label{eq:vmu}
\end{equation}
that transforms homogeneously under the chiral group as
\begin{equation}
V_{\mu} \, \mapright{G} \, h(g,\varphi) \, V_{\mu} \, h(g,\varphi)^{\dagger}
\label{eq:vmhvh}
\end{equation}
and ideal mixing, i.e. $V_8^{\mu} \, = \, ( \, \omega^{\mu} \, + \, \sqrt{2} \, 
\phi^{\mu} \, ) \, / \, \sqrt{3}$, is assumed.
\par
The most general strong vector--pseudoscalar--photon ($VP\gamma$) vertex,
assuming nonet symmetry, at leading ${\cal O}(p^3)$ reads \cite{EP90}
\begin{equation}
{\cal L}(VP\gamma) \, = \, h_V \, \varepsilon_{\mu \nu \rho \sigma}
 \, \langle \, V^{\mu} \, \{ \, u^{\nu} \, , \, f_+^{\rho \sigma} \, \}
\rangle \; \; \; \; , 
\label{eq:vpga}
\end{equation}
where
\begin{eqnarray}
f_{+}^{\mu \nu} \, = \, u \, F_L^{\mu \nu} \, u^{\dagger} \, + \, 
                        u^{\dagger} \, F_R^{\mu \nu} \, u \; \; \; 
& , & \; \; f_{+}^{\mu \nu} \, \mapright{G} \, 
h(g,\varphi) \, f_{+}^{\mu \nu} \, h(g,\varphi)^{\dagger} \; \; \; , 
\label{eq:fplus}
\end{eqnarray}
and $F_{R,L}^{\mu \nu}$ are the strength field tensors associated to the
external $r_{\mu}$ and $\ell_{\mu}$ fields. In Eq.~(\ref{eq:vpga}), from
the experimental width $\Gamma \, (\omega \rightarrow \pi^0 \gamma)$ 
\cite{PDG94}, $|h_V| \, = \, (3.7 \, \pm 0.3) \, \times \, 10^{-2}$.
\par
The most general vector--photon ($V\gamma$) coupling, at leading 
${\cal O}(p^3)$, can be written as
\begin{equation}
{\cal L}(V\gamma) \, = \, - \, \Frac{f_V}{2 \sqrt{2}} \, \langle
V_{\mu \nu} f_+^{\mu \nu} \, \rangle \; \; \; \; ,
\label{eq:fvvg}
\end{equation}
where $V_{\mu \nu} \, = \, \nabla_{\mu} V_{\nu} \, - \, 
\nabla_{\nu} V_{\mu}$ and $\nabla_{\mu}$ is the covariant derivative 
defined in \cite{EG89} as
\begin{eqnarray}
\nabla_{\mu} \, A \, & = & \, \partial_{\mu} \, A \, + \, 
\left[ \Gamma_{\mu} \, , \, A \, \right] \nonumber \\
& & \label{eq:cova} \\
\Gamma_{\mu} \, & \equiv & \, \Frac{1}{2} \, 
\left\{ \, u^{\dagger} \, ( \partial_{\mu} \, - \, i \, r_{\mu} \, ) \, u 
\, + \, u \, ( \partial_{\mu} \, - \, i \, \ell_{\mu} \, ) \, u^{\dagger}
\, \right\} \nonumber
\end{eqnarray}
for any $A$ operator that transforms homogeneously as the vector field in 
Eq.~(\ref{eq:vmhvh}). In Eq.~(\ref{eq:fvvg}), $|f_V| \, \simeq \, 0.20$
as obtained from the experimental width $\Gamma \, ( \rho^0 \rightarrow
e^+ e^-)$ \cite{PDG94}.
Moreover the positive slope of the $\pi^{0} \rightarrow
\gamma \gamma^*$ form factor determined experimentally tells us that the
effective couplings in Eqs.~(\ref{eq:vpga}, \ref{eq:fvvg}) must satisfy
$h_V f_V \, > \, 0$. 
\par
It has to be noted that the incorporation of vector mesons in chiral 
lagrangians is not unique and several realizations of the vector field
can be employed. 
In particular the antisymmetric formulation of vector fields was seen
to implement automatically vector meson dominance at \opc in $\chi$PT 
\cite{EG89}.
In \cite{EGL89} was shown that, at \opc in $\chi$PT,
once high energy QCD constraints are taken into account, the usual
realizations (antisymmetric, vector, Yang--Mills and Hidden formulations)
are equivalent.
 Although the antisymmetric tensor formulation of
spin--1 mesons was proven to have a better high--energy behaviour than
the vector field realization at \opc, this fact is not necessarily
the case in general. In fact for the odd--intrinsic parity coupling 
relevant in $V \rightarrow P \gamma$ decays, 
the antisymmetric tensor formulation
only contributes at ${\cal O}(p^8)$ while QCD requires explicit
\ops terms \cite{EP90}. This is the reason why we will use the vector field 
formulation (already introduced in Eqs.~(\ref{eq:vpga},\ref{eq:fvvg}))
that provides the right behaviour in the processes of our interest. We
will also see that the hidden gauge formalism \cite{BK88} provides also
the straightforward \ops operators.
\par
For a further extensive and thorough exposition on $\chi$PT see
\cite{DGH92,CGE95,TONI95}.

\subsection{Non--leptonic weak interactions in $\chi$PT}
\hspace*{0.5cm}
At low energies ($E \ll M_W$) the $\Delta S =1$ effective weak hamiltonian
(where only the degrees of freedom of the light quark fields remain) is
obtained from the Lagrangian of the Standard Model by using the asymptotic
freedom property of QCD in order to integrate out the fields with heavy
masses down to scales $\mu < m_c$. It reads
\begin{equation}
{\cal H}_{NL}^{|\Delta S| = 1} \; = \; 
- \, \Frac{G_F}{\sqrt{2}} \, V_{ud} V_{us}^* \, 
\sum_{i=1}^6 \, C_i(\mu) \, Q_i \; + \; h.c. \; \; \; \; . 
\label{eq:heffd}
\end{equation}
Here $G_F$ is the Fermi constant, $V_{ij}$ are elements of the 
CKM-matrix, $C_i(\mu)$ are the Wilson coefficients of the four--quark
operators $Q_i, \, i=1,...6$. In Eq.~(\ref{eq:heffd}) only 6 operators
appear if no virtual electromagnetic interactions nor leptons are 
included \cite{EDU95,CFM95,BUBU95}. 
In Eq.~(\ref{eq:heffd}) $Q_3, ..., Q_6$ are the QCD 
penguin operators \cite{VZS77}.
\par
If we neglect QCD corrections Eq.~(\ref{eq:heffd}) reduces to
\begin{eqnarray}
{\cal H}_{NL}^{|\Delta S|=1} \; & = & \; - \, \Frac{G_F}{\sqrt{2}} 
\, V_{ud} V_{us}^* \, Q_2 \; + \; h.c. \nonumber \\
& & \label{eq:noqcd} \\
& = & 
\; -  \, \Frac{G_F}{\sqrt{2}} \, 
V_{ud} V_{us}^* \, 4 \, ( \, \overline{s}_L \, \gamma^{\mu} \, u_L \, ) 
(\overline{u}_L \, \gamma_{\mu} \, d_L \, ) \; + \; h.c. \; \; , \nonumber 
\end{eqnarray}
with $\overline{s}_L \gamma_{\mu} u_L \, \equiv \, \frac{1}{2} 
\overline{s}^{\alpha} \gamma_{\mu} (1-\gamma_5) u_{\alpha}$, and
$\alpha$ a colour index.
\par
If the effect of gluon exchange at leading order is considered, the 
operator $Q_2$ in Eq.~(\ref{eq:noqcd}) is not multiplicatively 
renormalized but mixes with a new operator \cite{GAL74,AMA74}
\begin{equation}
Q_1 \, \equiv \, 4 \, (\, \overline{s}_L \, \gamma^{\mu} \, d_L \, ) 
(\, \overline{u}_L \, \gamma_{\mu} \, u_L \, ) \; \; \; \; .
\label{eq:q1only}
\end{equation}
Then we can consider the linear combinations
\begin{equation}
Q_{\pm} \, = \, Q_2 \, \pm \, Q_1 \; \; \; \; ,
\label{eq:qplusm}
\end{equation}
that do not mix under renormalization. 
Therefore taking into account only $Q_1$ and $Q_2$ we can write
\begin{equation}
{\cal H}_{NL}^{|\Delta S|=1} \, = \, - \, 
\Frac{G_F}{2 \sqrt{2}} \, V_{ud} V_{us}^* \, \left[ \, 
C_{+}(\mu) \, Q_{+} \, + \, C_{-}(\mu) \, Q_{-} \, \right]
\; + \; h.c. \; \; \; . 
\label{eq:hds1p}
\end{equation}
Considering the isospin quantum numbers  $Q_{+} \, \sim \, \Delta I
\, = \, 3/2$ and $Q_{-} \, \sim \, \Delta I \, = \, 1/2$. Already the
leading order computation of the Wilson coefficients shows that the 
$C_{+}$ coefficient is suppressed and the $C_{-}$ enhanced indicating
the right direction addressed by the phenomenology. However quantitatively
the enhancement is not big enough. In our further discussion we will 
not take into account the $\Delta I = 3/2$ transitions.
\par
The $Q_{-}$ operator 
transforms under $SU(3)_L \otimes SU(3)_R$ as the $( 8_L \, , \, 1_R )$
representation.
Then at ${\cal O}(p^2)$ we can construct
the effective lagrangian using the left--handed currents associated
to the chiral transformations, 
\begin{eqnarray}
{\cal L}_2^{|\Delta S|=1} \, & = & \, 4 \, \Frac{G_F}{\sqrt{2}} \, 
V_{ud} V_{us}^* \, g_8 \, \langle \, 
\lambda_6 \, L_1^{\mu} \, L^1_{\mu} \, \rangle \; \nonumber \\
& & \label{eq:weakla} \\
& = & \, \Frac{G_F}{\sqrt{2}} \, F^4 \, V_{ud} V_{us}^* \, g_8 \, 
\langle \, \Delta \, u_{\mu} \, u^{\mu} \, 
\rangle \; \; \; \; , \nonumber 
\end{eqnarray}
where 
\begin{equation} 
L_{\mu}^1 \, = \, \Frac{\delta S_2^{\chi}}{\delta \ell^{\mu}} \; = \; 
- \, i \, \Frac{F^2}{2} \, U^{\dagger} \, D_{\mu} \, U \; = \;  
- \, \Frac{F^2}{2} u^{\dagger} \, u_{\mu} \, u \; \; \; \; \; , 
\label{eq:leftcu}
\end{equation}
is the left--handed current associated to the $S_2^{\chi}$ action of 
the ${\cal O}(p^2)$ strong lagrangian Eq.~(\ref{eq:str2}) and 
$\Delta \, = \, u \, \lambda_6 \, u^{\dagger}$.
From the experimental width of 
$K \rightarrow \pi \pi$
and the use of ${\cal L}_2^{|\Delta S|=1}$  Eq.~(\ref{eq:weakla}), 
that is, at ${\cal O}(p^2)$ one gets \footnote{If \opc corrections are taken
into account, a phenomenological value of $|g_8| \simeq 3.6$ is obtained
\cite{KMW91}.} 
\begin{equation}
|g_8|_{K \rightarrow \pi \pi} \, \simeq \, 5.1 \; \; \; \; \; \; , 
\; \; \; \; \; \; 
G_8 \, \equiv \, \Frac{G_F}{\sqrt{2}} \, V_{ud} V_{us}^* \, 
|g_8|_{K \rightarrow \pi \pi}  \, \simeq \, 9.2 \, 
\times \, 10^{-6} \, GeV^{-2} \; \; \; . 
\label{eq:g8fr}
\end{equation}
If instead we use the result for 
the Wilson coefficient $C_{-}(m_{\rho})$ at leading ${\cal O}(\alpha_s)$
and with $\Lambda_{\overline{MS}} \, = \, 325 \, MeV$ we get 
\begin{equation}
g_8^{Wilson} \, = \, \Frac{1}{2} C_{-} (m_{\rho}) \;  
\simeq \; 1.1 \; \; \; \; .
\label{eq:g81}
\end{equation}
\par
At \opc the chiral weak lagrangian has been studied in \cite{KM90,EK93} giving
37 chiral operators $W_i$ only in the octet part
\begin{equation}
{\cal L}_4^{|\Delta S|=1} \, = \, G_8 \, F^2 \, 
\sum_{i=1}^{37} \, N_i \, W_i \; + \; h.c. \; \; \; \; \; , 
\label{eq:weak4}
\end{equation}
that gives 37 new coupling constants $N_i$. Their phenomenological
determination is then very difficult and models are necessary in order to 
predict them \cite{EK93,IP92}. 

\section{\kpiggtot and \kggs amplitudes}
\hspace*{0.5cm}
The general amplitude for \kpiggtot is given by
\begin{equation}
M\, ( \, K (p) \rightarrow \pi  \, \gamma (q_1,\epsilon_1) \, 
\gamma (q_2,\epsilon_2) \, ) \; = \; {\epsilon_1}_{\mu} {\epsilon_2}_{\nu}
\, M^{\mu \nu} (p,q_1,q_2) \; \; \; , 
\label{eq:mee}
\end{equation}
where $\epsilon_1$,$\epsilon_2$ are the photon polarizations, and 
$M^{\mu \nu}$ has four invariant amplitudes \cite{EPR88}
\begin{eqnarray}
M^{\mu \nu}  & = & \Frac{A(z,y)}{m_K^2} \, ( q_2^{\mu} \,  q_1^{\nu} \, - \, 
q_1 \cdot q_2 g^{\mu \nu} ) \;
 + \; \Frac{2\, B(z,y)}{m_K^4} \, ( -p \cdot q_1 \, p \cdot q_2 \, 
g^{\mu \nu}
\, - \, q_1 \cdot q_2 \, p^{\mu} p^{\nu}  \, \nonumber \\
& & \qquad \qquad \qquad \qquad \qquad \qquad \qquad \qquad \qquad \; 
+ \, p \cdot q_1 \, q_2^{\mu} p^{\nu} 
\, + \, p \cdot q_2 \, p^{\mu} q_1^{\nu} \, \,  )  \nonumber \\
& & \label{eq:mmunu} \\
& & + \; \Frac{C(z,y)}{m_K^2} \; \varepsilon^{\mu \nu \rho \sigma} 
{q_1}_{\rho} {q_2}_{\sigma} \,  
 + \; \Frac{D(z,y)}{m_K^4} \; [ \; \varepsilon^{\mu \nu \rho \sigma}
\, ( p \cdot q_2 \, {q_1}_{\rho} + p \cdot q_1 \, {q_2}_{\rho} ) p_{\sigma}
\nonumber \\
& & \qquad \qquad \qquad \qquad \qquad \qquad \qquad \; \; \, + \, 
( p^{\mu} \varepsilon^{\nu \alpha \beta \gamma} \, + \, p^{\nu} 
\varepsilon^{\mu \alpha \beta \gamma} ) \, p_{\alpha} {q_1}_{\beta} 
{q_2}_{\gamma} \; ] \nonumber
\end{eqnarray}
and
\begin{eqnarray}
y \, = \, \Frac{p \cdot (q_1 - q_2)}{m_K^2} \, & \; \; \; \;  , & \;  
\; \; \; \; z \, = \, \Frac{(q_1 + q_2)^2}{m_K^2}  \; \; \; .
\label{eq:yz}
\end{eqnarray}
The physical region in the adimensional variables $y$ and $z$ is given
by : 
\begin{eqnarray}
0 \, \leq \, |y| \, \leq \, \Frac{1}{2} \lambda^{1/2} \,(1,r_{\pi}^2,z) \; & \; 
\; , \; \; & \; 0 \, \leq \, z \leq \, (1-r_{\pi})^2 \; \; \; \; ,
\label{eq:range}
\end{eqnarray}
with $r_{\pi} = m_{\pi}/m_K$ and 
$\lambda (a,b,c)  =  a^2  +  b^2  +  c^2  -  2  
( a b  +  a c  +  b c  )  \, . $
Note that the invariant amplitudes $A(z,y)$, $B(z,y)$ and $C(z,y)$ have
to be symmetric under the interchange of $q_1$ and $q_2$ as required
by Bose symmetry, while $D(z,y)$ is antisymmetric. In the limit where 
CP is conserved the amplitudes $A$ and $B$ contribute only to \kpiggn
while $C$ and $D$ only contribute to $K_S \rightarrow \pi^{0}
\gamma \gamma$. In \kpiggp all of them are involved. 
\par
Using the definitions Eqs.~(\ref{eq:mmunu},\ref{eq:yz}) the double differential
rate for unpolarized photons is given by 
\begin{eqnarray}
\Frac{\partial^2 \Gamma}{\partial y \, \partial z} & \, = \, & 
\, \Frac{m_K}{2^9 \pi^3} \left[ \, z^2 \left( \, | \, A \, + \, B \, |^2
 \, + \, | \, C \, |^2 \right)  \; \right. \label{eq:doudif} \\
& & \; \; \;  \; \; \; \; \; \; \; \left.
+ \, \left( \, y^2 \, - \, \Frac{1}{4} \lambda (1,r_{\pi}^2,z) \, \right)^2
\, \left( \, | \, B \, |^2 \, + \, | \, D \, |^2 \right) \, \right]
\; \; \; . \nonumber
\end{eqnarray}
The processes $K \rightarrow \pi \gamma \gamma$ have no tree level
${\cal O}(p^2)$ contribution because there are not enough powers of 
momenta to satisfy the constraint of gauge invariance. For this same
reason at ${\cal O}(p^4)$ the amplitudes
$B$ and $D$ are still zero. Therefore their leading
contribution is ${\cal O}(p^6)$. As can be seen from Eq.~(\ref{eq:doudif})
only the $B$ and $D$ terms contribute for small $z$ (the invariant
amplitudes are regular in the small $y,z$ region). The antisymmetric
character of the $D$ amplitude under the interchange of $q_1$ 
and $q_2$ means effectively that while its leading contribution is 
${\cal O}(p^6)$ this only can come from a finite loop calculation
because the leading counterterms for the $D$ amplitude are 
${\cal O}(p^8)$ \footnote{This is so due to the fact that its antisymmetric 
property demands that its leading contribution cannot be a constant
term.}. However also this loop contribution is helicity suppressed
compared to the $B$ term. As shown in a similar situation in the
electric Direct Emission of $K_L \rightarrow \pi^+ \pi^- \gamma$
\cite{DI95} this antisymmetric ${\cal O}(p^6)$ loop contribution might
be smaller than the local ${\cal O}(p^8)$ contribution. 
\par
Thus in the 
region of small $z$ (collinear photons) the $B$ amplitude is dominant
and can be determined separately from the $A$ amplitude. This feature
is important in order to evaluate the CP conserving contribution
$K_L \rightarrow \pi^0 \gamma \gamma \rightarrow \pi^0 e^+ e^-$. Both
on--shell and off--shell two--photon intermediate states generate, through
the $A$ amplitude, a contribution to $K_L \rightarrow \pi^0 e^+ e^-$
that is proportional to $m_e/m_K$ and therefore suppressed \cite{EPR88}. 
Otherwise
the $B$--type amplitude, though appearing only at \ops, generates 
 a relevant unsuppressed 
contribution to $K_L \rightarrow \pi^0 e^+ e^-$ through the
on--shell photons \cite{DH87,SE88,FR89}, due to the different helicity
structure.
The dispersive two--photon contribution generated by the 
$B$--type amplitude has been approximated by choosing an appropriate
form factor for the virtual photon coupling \cite{DG95}.
\par
The amplitude for \kggs (neglecting any CP--violating effects) is given
by
\begin{equation}
M (\, K_L(p) \rightarrow \gamma (q_1, \epsilon_1) \gamma^* 
(q_2, \epsilon_2) \, )
\; = \; i \, A_{\gamma \gamma^*} (q_2^2) \, \varepsilon_{\mu \nu \sigma \tau}
\, \epsilon_1^{\mu}(q_1) \epsilon_2^{\nu}(q_2) q_1^{\sigma} q_2^{\tau}
\; \; \; , 
\label{eq:klggs}
\end{equation}
The unique amplitude $A_{\gamma \gamma^*} (q_2^2)$ we can express as 
\begin{equation}
A_{\gamma \gamma^*} (q_2^2) \; = \;  \, A_{\gamma \gamma}^{exp} 
\; f(x) \; \; \; \; , 
\label{eq:aggs}
\end{equation}
where $A_{\gamma \gamma}^{exp}$ is the experimental amplitude 
$A(K_L \rightarrow \gamma \gamma)$ and $x=q_2^2/m_K^2$. The form factor
$f(x)$ in Eq.~(\ref{eq:aggs}) is properly normalized to $f(0)=1$. We define
the slope $b$ of $f(x)$ as 
\begin{equation}
f(x) \, = \, 1 \, + \, b \, x \, + \, {\cal O}(x^2) \; \; \; \; . 
\label{eq:bslope}
\end{equation}
The differential decay spectrum for \kgll, in absence of radiative
corrections, is given by 
\begin{equation}
\Frac{1}{\Gamma_{\gamma \gamma}} \Frac{d \, \Gamma}{d \, x} \; = \; 
\Frac{2 \alpha}{3 \pi} \, \Frac{(1-x)^3}{x} \, \left( \, 1 \, + \, 
\Frac{2 m_{\ell}^2}{x m_K^2} \, \right) \, 
\sqrt{ \, 1 \, - \, \Frac{4 m_{\ell}^2}{x m_K^2}} \, | \, f(x) \, |^2
\; \; \; , 
\label{eq:difra}
\end{equation}
where $\Gamma_{\gamma \gamma}$ is the $K_L \rightarrow \gamma \gamma$
decay rate. 

\section{Local contributions to \kpiggtot and \kgll generated by vector 
resonance exchange}
\hspace*{0.5cm}
The leading finite \opc amplitudes of \kpiggn were evaluated some time ago
\cite{DE87}, generating only the $A$--type amplitude in Eq.~(\ref{eq:mmunu}).
No local contributions arise since all the external particles 
involved are electrically neutral. Thus the loop amplitude is simply
predicted in terms of $G_8$ from ${\cal O}(p^2)$ $K \rightarrow \pi \pi$
Eq.~(\ref{eq:g8fr}) (and the corresponding $G_{27}$ when $\Delta I =3/2$
transitions are taken into account \cite{CD93}). 
\par
Also the leading \opc amplitudes for \kpiggp were computed in \cite{EPR88}.
At this order both the $A$-- and $C$--type amplitudes in Eq.~(\ref{eq:mmunu})
appear. The $C$ amplitude is generated by the Wess--Zumino--Witten anomalous
vertices $\pi^0, \eta \rightarrow \gamma \gamma$ \cite{WZW71} and consequently
is completely determined at leading \opc. Though the leading order is
finite, chiral symmetry allows a scale independent local contribution to the
$A$ amplitude. The counterterm combination can be written in terms of 
strong and weak effective couplings as
\begin{equation}
\hat{c} \; = \; \Frac{128 \, \pi^2}{3} \, \left[ \, 3 ( L_9 + L_{10}) \, 
+ \, N_{14} \, - \, N_{15} \, - \, 2 \, N_{18} \, \right] \; \; \; ,
\label{eq:chat}
\end{equation}
where $L_9$ and $L_{10}$ are known couplings of the \opc $\chi$PT 
strong lagrangian \cite{GL85} and $N_i$ are couplings of the 
\opc weak $\chi$PT lagrangian Eq.~(\ref{eq:weak4}) \cite{KM90,EK93}. 
Since weak counterterms are involved in Eq.~(\ref{eq:chat}), resonance
exchange is not sufficient to make predictions and models have to be 
invoked \cite{EK93,IP92,BP93}. The combination
of weak couplings in Eq.~(\ref{eq:chat}) is an independent relation
to take into account in order to determine the \opc weak couplings
\cite{DE95} and to test the models.
\par
The observed branching ratio for \kpiggn is 
\begin{eqnarray}
Br (\kpiggne) \, |_{exp} \, = \, (1.7 \pm 0.3) \times 10^{-6} \; \; \; 
\; \; \; \; \; \; & 
& \; \; \; \; \; [NA31, \cite{NA31}] \; \; \; , \nonumber \\
& & \label{eq:brklexp} \\
Br (\kpiggne) \, |_{exp} \, = \, (1.86 \pm 0.60 \pm 0.60) \times 10^{-6} 
\; & & \; \; \; \; \;  [E731, \cite{E731}] \; \; \; \; . \nonumber
\end{eqnarray}
which is substantially larger than the \opc prediction \cite{CD93,DE87}
$
Br (\kpiggne) \, |^{{\cal O}(p^4)}_{{\bf 8 + 27}} \; \simeq \; 0.61 \, \times
\, 10^{-6} \, . 
$
However the \opc spectrum of the diphoton invariant mass nearly agrees with
the experiment, in particular no events for small $m_{\gamma \gamma}$ are
observed, implying a small $B$--type amplitude. Thus \ops corrections have
to be important. Though no complete calculation is available, the 
supposedly larger contributions have been performed~: \ops unitarity 
corrections from $K_L \rightarrow \pi^0 \pi^+ \pi^-$ were worked out
in \cite{CD93,CE93}, they enhance the \opc branching ratio by $30 \%$,
\begin{equation}
Br (\kpiggne) \, |^{{\cal O}(p^6)}_{unitarity} \; \simeq \; 
0.84 \, \times \, 10^{-6} \; \; \; \; , 
\label{eq:brklthe}
\end{equation}
and generate a $B$--type amplitude slightly spoiling the satisfactory
\opc spectrum.
\par
\ops loop contributions to \kpiggtot are in general divergent and need
to be regularized.
Lorentz and gauge invariance give
four different local structures at \ops for \kpiggtot
processes~:
\begin{equation}
F_{\mu \nu} \, F^{\mu \nu} \, \partial_{\lambda} K \, 
\partial^{\lambda} \pi \; \; \;  ,  \; \; \;
m_K^2 \, F_{\mu \nu} F^{\mu \nu} \, K \, \pi \; \; \;  , \; \; \; 
\partial^{\alpha} F_{\mu \nu} \, \partial_{\alpha} F^{\mu \nu} \, 
K^+ \pi^-
 \; \; \; ,  \; \; \;
F_{\mu \nu} \, F^{\mu \lambda} \, \partial^{\nu} K \, \partial_{\lambda}
\pi \; . 
\label{eq:kind}
\end{equation}
Here both neutral and charged mesons are understood and the term with 
derivatives on the photon strength field tensor only appears for 
electrically charged mesons. 
It is necessary to emphasize that there is no reason why the effective
couplings of these structures have to coincide in the neutral and charged
channels because different chiral structures in both channels can give
the local terms in Eq.~(\ref{eq:kind}).
Between the different structures in 
Eq.~(\ref{eq:kind}) only the last one contributes to a $B$ amplitude in
Eq.~(\ref{eq:mmunu}).
\par
The physics behind the couplings weighting the weak local amplitudes in 
Eq.~(\ref{eq:kind}) has not been fully investigated. 
It is sound to think, however, that the lightest non--Goldstone meson 
spectrum plays a major r\^ole. In particular we will concentrate in the 
possible contribution of vector resonances and we disregard heavier states
which contribution is expected to be smaller.
In the generation of the
structures in Eq.~(\ref{eq:kind}) we find that only the terms with derivatives
on the meson fields can be generated by vector meson exchange.
\par
In \cite{EP90} two different vector generated amplitudes
 for \kpiggtot were considered : i) a strong vector exchange with a weak 
transition in a external leg (diagrams in Figs.~1.a and 1.b), and ii) a 
direct vector exchange between a weak $VP\gamma$ vertex and a strong one
(diagrams in Figs.~1.c and 1.d). The BMS model \cite{BMS83} suggests a 
third direct contribution: iii) a weak vector--vector transition as 
in diagram of Fig.~1.e.
While the first
contribution is well under control from the phenomenology of strong
interactions, the second case depends strongly on the model used for 
the weak $VP\gamma$ vertex. 
\par
The diagram in Fig.~1.a was advocated to generate large local contributions
to \kpiggn, such that, if added to the \opc loop amplitude, could
accomodate the disagreement in the width \cite{SE88,SH92}. However due
to the presence of a local $B$--type amplitude, the spectrum, particularly
at low $m_{\gamma \gamma}$, does not agree with experiment. Furthermore
in some models (like WDM) \cite{EP90} there is a tendency to a 
cancellation between diagrams in Figs.~1.a, 1.b and Figs.~1.c, 1.d. 
Nevertheless no definitive statement about the resonance exchange 
r\^ole has yet been made.
Cohen, Ecker and Pich noticed \cite{CE93} that by adding the 
contributions from the diagrams in Figs.~1.a,1.b,1.c and 1.d, for a choice
of the unknown weak coupling in diagrams Figs.~1.c and 1.d, one could, 
simultaneously, obtain the experimental spectrum and width of 
\kpiggn.
\par
The \ops unitarity corrections from $K^+ \rightarrow \pi^+ \pi^+ \pi^-$
to \kpiggp have been computed in \cite{DP96} and found to produce
a $B$--type amplitude. This contribution increases the rate by a 
$30$--$40 \%$. Differently from \kpiggn, \ops vector meson exchange seems
to be negligible in \kpiggp \cite{DP96}, at least in the FM and WDM.
It is worth stressing that the question of the size of \ops vector
meson exchange contribution can be studied independently from the $A$
amplitude in the region of small diphoton invariant mass. Thus the 
value of $\hat{c}$ in Eq.~(\ref{eq:chat}) can be studied in the
remaining kinematical region. 

\begin{figure}
\begin{center}
\leavevmode
\hbox{%
\epsfxsize=16cm
\epsffile{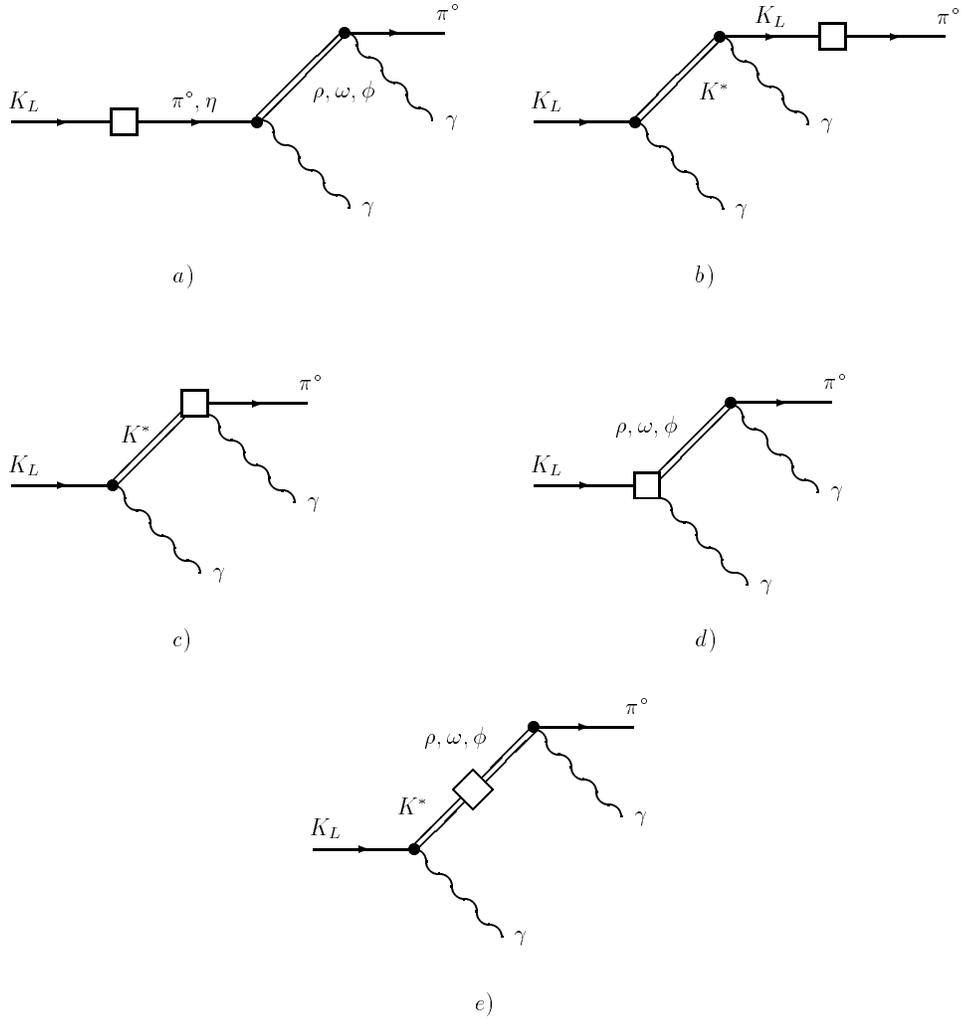}}
\end{center}
\vspace*{-6cm}
\caption{Feynman diagrams for \kpiggn . a) and b) correspond
to {\em external} weak transitions, c) and d) to {\em direct} weak
transitions. e) is the contribution of the BMS model to the {\em direct}
weak transitions. Analogous to these contribute to \kpiggp.}
\end{figure}

\subsection{The weak $VP\gamma$ vertex}
\hspace*{0.5cm}
The prospectives to a parallel analysis of \kpiggn and 
\kpiggp are very good and therefore the questions of the size of 
vector meson exchange and the completeness of the computed \ops
contributions can be answered. The situation will be even better if 
we know more accurately
the vector meson exchange contribution~: in this paper we pursue the 
idea that a good weak vector meson model has to describe simultaneously
the \ops slope in \kggs and the \ops vector meson contribution to 
\kpiggn ; then, as a further prediction, to \kpiggp.
\par
Hence we can think of a general framework where we consider the 
full structure
of the weak $VP\gamma$ vertex that belongs to the octet representation
of $SU(3)_L$ \footnote{We will not consider operators of the {\bf 27}
representation of $SU(3)_L$ because their contribution is presumably
much less important.}. Then the most general effective weak coupling
$VP\gamma$ able to contribute to \ops \kpiggtot and \kggs processes is 
\begin{equation}
{\cal L}_W(VP\gamma) \; = \; G_8 \, F_{\pi}^2 \, \langle \, V^{\mu}
\, {\cal J}_{\mu}^W \, \rangle \; \; \; \; ,
\label{eq:mostg}
\end{equation}
where
\begin{equation}
{\cal J}_{\mu}^W \; = \; \varepsilon_{\mu \nu \alpha \beta} \; 
\sum_{i=1}^{5} \, \kappa_i \, T_i^{\nu \alpha \beta} \; \; \; ,
\label{eq:fullcur}
\end{equation}
and
\begin{eqnarray}
T_1^{\nu \alpha \beta} \; & = & \;  \{ \, u^{\nu} \, , \, \Delta \, 
                        f_{+}^{\alpha \beta} \, \} \; \; \; \; , 
\nonumber  \\
T_2^{\nu \alpha \beta} \; & = & \;  
          \{ \, \{ \Delta \, , \, u^{\nu} \, \} \, , \, 
                         f_{+}^{\alpha \beta} \, \} \; \; \; \; ,  
\nonumber \\
T_3^{\nu \alpha \beta} \; & = & \; 
      \langle \, u^{\nu} \, \Delta \, \rangle \, 
                        f_{+}^{\alpha \beta}   \; \; \; \; , 
\nonumber \\
T_4^{\nu \alpha \beta} \; & = & \; 
      \langle \, u^{\nu} \, f_{+}^{\alpha \beta} \, 
                      \rangle \, \Delta  \; \; \; \; , 
\nonumber  \\
T_5^{\nu \alpha \beta} \; & = & \; 
     \langle \, \Delta \, u^{\nu} \, f_{+}^{\alpha \beta} \, \rangle 
\; \; \; \; .
\label{eq:fullcurr1}
\end{eqnarray}
In Eq.~(\ref{eq:fullcur}) $\kappa_i$, $i=1,2,3,4,5$ are dimensionless
coupling constants to be determined from phenomenology or theoretical
models.
There are other terms for processes involving more pseudoscalars. We use,
as anticipated in Section 2.1, the usual formulation of vector fields to
describe the vector mesons.

\subsection{\kpiggtot}
\hspace*{0.5cm}
By integrating out the vector meson fields interchanged between
one vertex from ${\cal L}_W(VP\gamma)$ Eq.~(\ref{eq:mostg}) and a 
vertex from ${\cal L}(VP\gamma)$ Eq.~(\ref{eq:vpga}) we generate a lagrangian
that provides local \ops contributions to \kpiggtot. The result is
\begin{eqnarray}
{\cal L}_6^W \, & = & - \, \Frac{128 \pi}{9} \, G_8 \, \alpha_{em} \, 
\Frac{h_V}{m_V^2} \, 
\left[ \, F^{\mu \nu} \, F_{\mu \nu} \, 
          \left[ \, ( 2 \kappa_1 \, + \, 4 \kappa_2 \, - \, 3 \kappa_3 \, 
                      + \, 3 \kappa_4
                      \, + \, 3 \kappa_5 ) \, \partial^{\alpha} K_2^{0}
                     \, \partial_{\alpha} \pi^{0} \,
\right. \right. \nonumber \\
& & \qquad \qquad \qquad \qquad \qquad \qquad \; \; \;  
\left. \left.  + \,   (\kappa_1 \, - \, \kappa_2) \, ( \partial^{\alpha} K^+
                  \,   \partial_{\alpha} \pi^- \, + \, \partial^{\alpha} K^-
                  \,   \partial_{\alpha} \pi^+ \, ) \, 
           \right] \right. \nonumber \\
& & \label{eq:l6wvec} \\
& & \; \; \; \; \; \; \; \; \; \; \; \; \; \; \; \; \; \; \; 
\qquad \left. + \, 2 \, 
F^{\mu \nu} \, 
F_{\nu \lambda} \, 
          \left[ \, ( 2 \kappa_1 \, + \, 4 \kappa_2 \, - \, 3 \kappa_3 \, 
                      + \, 3 \kappa_4
                      \, + \, 3 \kappa_5 ) \, \partial^{\lambda} K_2^{0}
                     \, \partial_{\mu} \pi^{0} \, 
\right. \right. \nonumber \\
& & \qquad \qquad \qquad \qquad \qquad \qquad \; \; \; \; 
\left. \left.
             + \,   (\kappa_1 \, - \, \kappa_2) \, ( \partial^{\lambda} K^+
                  \,   \partial_{\mu} \pi^- \, + \, \partial^{\lambda} K^-
                  \,   \partial_{\mu} \pi^+ \, ) \, 
           \right] 
\right] \; \; \; \; , \nonumber 
\end{eqnarray}
where $m_V$ is the degenerate mass of the vector mesons in the chiral limit
and $K^0_2 = K_L$ in the CP limit. 
\par
If we compare Eq.~(\ref{eq:l6wvec}) with Eq.~(\ref{eq:kind}) we see that, as we
already commented, only the terms with derivatives over the meson fields
are generated by vector resonance exchange. ${\cal L}_6^W$ generates
$A$ and $B$ amplitudes in Eq.~(\ref{eq:mmunu}). Moreover Vector Meson 
Dominance and the restrictions of chiral symmetry show that the 
combination of weak structures in Eq.~(\ref{eq:fullcur}) to 
$F^{\mu \nu} F_{\mu \nu} \partial^{\alpha} K \partial_{\alpha} \pi$ 
and 
$F^{\mu \nu} F_{\nu \lambda} \partial^{\lambda} K \partial_{\mu} \pi$ 
in Eq.~(\ref{eq:l6wvec}) is the same.
\par
Following \cite{EP90} we define the local contribution generated by 
vector resonance exchange $a_V$ as
\begin{equation}
a_V \, = \, - \, \Frac{\pi}{2 G_8 m_K^2 \alpha_{em}} \, 
\lim_{z \rightarrow 0} \, B_V(z) \; \; \; \; ,
\label{eq:avb}
\end{equation}
where $B_V(z)$ is the vector resonance contribution to the $B$ amplitude. 
In \cite{EP90} two sources for $a_V$ were discussed~: 
\begin{equation}
a_V \, = \, a_V^{ext} \, + \, a_V^{dir} \; \; \; , 
\label{eq:avsplit}
\end{equation}
that is 
i) strong
vector resonance exchange with an external weak transition ($a_V^{ext}$)
as in Fig.~1.a and Fig.~1.b, 
ii) direct vector resonance exchange between a weak and a strong 
$VP\gamma$ vertices ($a_V^{dir}$) as in Fig.~1.c and Fig.~1.d.
The BMS model suggests the existence of a further $a_V^{dir}$ contribution
as given by the diagram in Fig.~1.e.
\par
The contribution to $a_V$ from Vector Resonance exchange between two
strong vertices supplemented with a weak transition in an external leg
(Figs.~1.a and 1.b) is very well determined phenomenologically due to 
our good understanding of the strong sector involved. This we call
the {\em external} contribution ($a_V^{ext}$) and has been worked out
in \cite{EP90} (we write the subscript $0$ for the 
neutral channel \kpiggn~: $a_{V,0}$ ; and the subscript $+$ for 
the charged channel \kpiggp~: $a_{V,+}$)~:
\begin{eqnarray}
a_{V,0}^{ext} \; & = & \; \Frac{512 \pi^2}{9} \, h_V^2 \, 
\Frac{m_K^2}{m_V^2} \, \,  \simeq \; 0.32 \; \; \; , \nonumber \\
& & \label{eq:avextc} \\
a_{V,+}^{ext} \; & = & \; - \, \Frac{a_{V,0}^{ext}}{4} \, \,  
\simeq \; -0.08 \; \; \; , \nonumber 
\end{eqnarray} 
where $h_V$ is given in Eq.~(\ref{eq:vpga}) and we have used $m_V \, = \, 
m_{\rho}$ for the numerical evaluation.
\par
The second contribution ($a_V^{dir}$) depends on 
the model for the direct $VP\gamma$ vertex, but the general structure
(contributing to the processes of our interest) is the one provided by 
${\cal L}_6^W$ Eq.~(\ref{eq:l6wvec}) with the following expression~:
\begin{eqnarray}
a_{V,0}^{dir} \; & = & \; - \, \Frac{128 \pi^2}{9} \, h_V \, 
\Frac{m_K^2}{m_V^2} \, \left[ \, 2 \kappa_1 \, + \, 4 \kappa_2 \, - 
\, 3 \kappa_3 \, + \, 
3 \kappa_4 \, + \, 3 \kappa_5 \, \right] \; \; \; , \nonumber \\
& & \label{eq:avdirk} \\
a_{V,+}^{dir} \; & = & \; - \, \Frac{128 \pi^2}{9} \, h_V \, 
\Frac{m_K^2}{m_V^2} \, \left[ \, \kappa_1 \, - \, \kappa_2 \, \right]
\; \; \; \; \; . \nonumber
\end{eqnarray}

\begin{figure}
\begin{center}
\leavevmode
\hbox{%
\epsfxsize=16cm
\epsffile{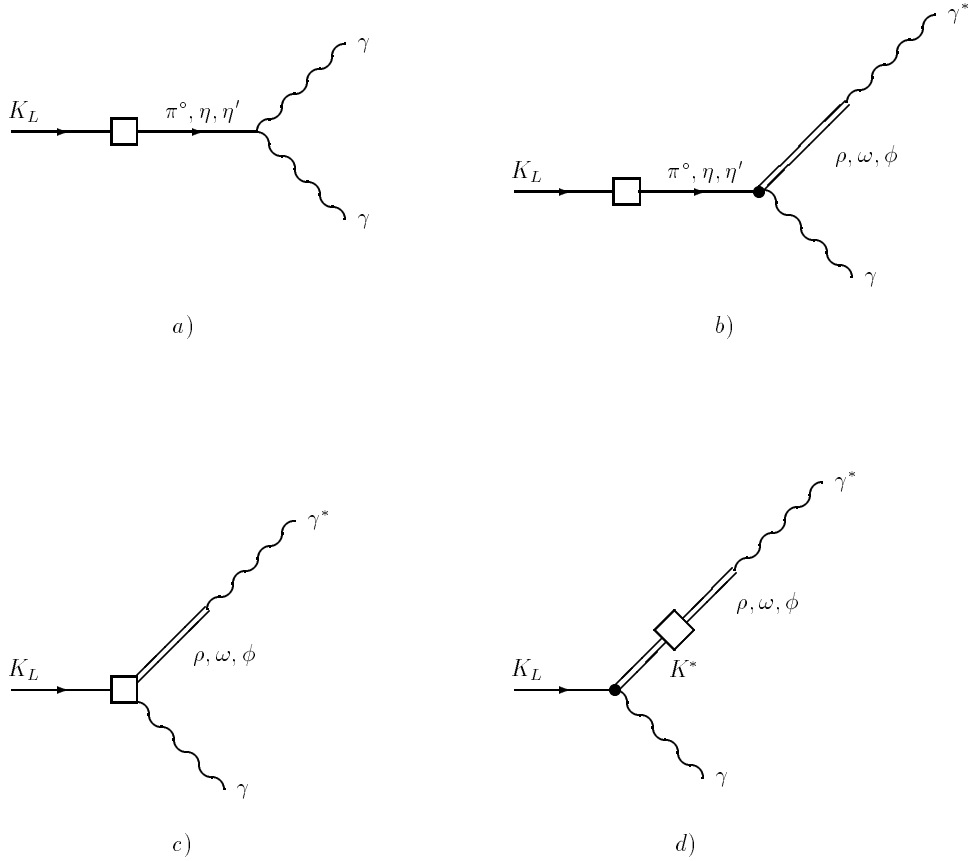}}
\end{center}
\vspace*{-7cm}
\caption{Feynman diagrams for \kggs. a) corresponds to the pole
model contribution to both
photons on--shell, b) represents the {\em external} weak transition, 
c) the {\em direct} weak transition and d) is the contribution of the
BMS model to the {\em direct} weak transition.}
\end{figure}

Cohen, Ecker and Pich \cite{CE93} have shown that unitarity 
corrections from 
$K_L \rightarrow \pi^0 \pi^+ \pi^-$ and vector meson exchange with 
$a_{V,0}\simeq -0.9$ restore the agreement with 
experiment. 
The comparative analysis with \kpiggp, which is going to be 
measured soon at BNL--787, should establish the dominance of the
computed \ops contributions (and, simultaneously, the value of 
$\hat{c}$ in Eq.~(\ref{eq:chat})) also for \kpiggn. Thus $a_{V,0}$ 
should be more safely determined.

\subsection{\kggs}
\hspace*{0.5cm}
We can also use our weak effective vertex $VP\gamma$ 
Eqs.~(\ref{eq:mostg}, \ref{eq:fullcur}) to evaluate its 
contribution to the slope
of the form factor $f(x)$ in Eqs.~(\ref{eq:aggs}, \ref{eq:bslope}). 
\par
The slope $b$ gets two different contributions:
\begin{equation}
b \, = \, b_V \, + \, b_D \; \; \; , 
\label{eq:bsplit}
\end{equation} 
i) the first one ($b_V$) 
comes from the strong vector interchange with the weak transition
in the $K_L$ leg as shown in Fig.~2.b, ii) a direct weak transition
$K_L \rightarrow V \gamma$ ($b_D$) as in Fig.~2.c. 
In \cite{BMS83} the direct transition was constructed as in Fig.~2.d, i.e.
with a weak vector--vector transition (BMS model).
\par
The evaluation of $b_V$ is straightforward \cite{ECK90} \footnote{ 
Ecker's definition $r_V$ relates with our $b_V$ through $b_V = r_V m_K^2 /
m_V^2$.}. By integrating out the vector mesons between two vertices
generated respectively by ${\cal L}(VP\gamma)$ Eq.~(\ref{eq:vpga})
and ${\cal L}(V\gamma)$ Eq.~(\ref{eq:fvvg}) and assuming nonet symmetry
in the pseudoscalar sector we get the effective lagrangian
\begin{equation}
{\cal L}_{VMD}^6 \; = \; \Frac{16 \sqrt{2}}{3} \pi \alpha_{em}
\Frac{h_V f_V}{F_{\pi} m_V^2} \varepsilon_{\nu \lambda \alpha \beta} \, 
F^{\alpha \beta} \, \partial_{\mu} F^{\mu \nu} \, 
\partial^{\lambda} \, \left( \, \pi^{0} \, + \, \Frac{\eta_8}{\sqrt{3}} 
\, + \, 2 \sqrt{\Frac{2}{3}} \eta_1 \, \right) \; \; \; . 
\label{eq:l6vmd}
\end{equation}
Since the slope is normalized by $A(K_L \rightarrow \gamma \gamma)$, this
lagrangian has to be compared with the one generated by 
Eq.~(\ref{eq:zulr})~:
\begin{equation}
{\cal L}_{WZW} \, = \, - \, \Frac{\alpha_{em}}{8 \pi F_{\pi}} \, 
\varepsilon_{\mu \nu \alpha \beta} \, F^{\mu \nu} \, F^{\alpha \beta} \, 
\left( \, \pi^0 \, + \, \Frac{\eta_8}{\sqrt{3}} \, + \, 
2 \sqrt{\Frac{2}{3}} \, \eta_1 \, \right) \; \; \; \; .
\label{eq:piog}
\end{equation}
Implementing the external weak transition $K_L \rightarrow \pi^{0},
\eta, \eta'$ as given by Eq.~(\ref{eq:weakla}) (extended to 
$U(3) \otimes U(3)$)
and normalizing with the amplitude $K_L \rightarrow \gamma
\gamma$ (Fig.~2.a) we get for $b_V$ a result independent from weak
interaction parameters
\begin{equation}
b^{nonet}_V \; = \; \Frac{32 \sqrt{2}}{3} \pi^2 f_V h_V \, 
\Frac{m_K^2}{m_V^2} \; 
\simeq \; 0.46 \; \; \; \; ,  
\label{eq:rvst}
\end{equation}
and the condition $h_V f_V > 0$ from $\pi^{0} \rightarrow \gamma \gamma^*$
gives $b_V > 0$. It has to be noted that this result for $b_V$ is entirely
due to the singlet $\eta_1$ contribution. This is so because the 
Gell-Mann--Okubo mass relation cancels the $\pi^0$ and $\eta_8$ contribution
and then 
\begin{equation}
b^{octet}_V \, = \, 0 \; \; \; \; .
\label{eq:bocte}
\end{equation}
Violations of the mass relation or contribution from the $\eta$-$\eta'$ 
mixing are at higher chiral order.
For consistency we think one should assume nonet or octet realizations
simultaneously in $b_D$ and $b_V$. 
\par
We can parameterize $b_D$ in the general framework using Eqs.~(\ref{eq:fvvg}, 
\ref{eq:mostg}) as
\begin{equation} 
b_{D} \;  =  \; - \, \Frac{64 \sqrt{2}}{9} \pi \, G_8 \, \alpha_{em}
\, F_{\pi} \, f_V \, \Frac{m_K^2}{m_V^2} \, 
\Frac{1}{|A_{\gamma \gamma}^{exp}|} \; 
\left[ \, \kappa_1 \, + \, 2 \kappa_2 \, + \, 3 \kappa_3 \, \right]
\; \; \; \; . \label{eq:bstar}
\end{equation}
By comparing the expression of $a_{V,0}^{dir}$ in Eq.~(\ref{eq:avdirk}) and 
$b_D$ in Eq.~(\ref{eq:bstar}) we can get a general relation between
both observables as 
\begin{equation}
\Frac{a^{dir}_{V,0}}{b_D} \, = \, \sqrt{2} \, \pi \, 
\Frac{h_V}{f_V} \, 
\Frac{|A_{\gamma \gamma}^{exp}|}{G_8 \, F_{\pi} \, \alpha_{em}}
\, \Frac{2 \kappa_1 + 4 \kappa_2 - 3 \kappa_3 + 3 \kappa_4 + 
3 \kappa_5}{\kappa_1 + 2 \kappa_2 + 3 \kappa_3} \; \; \; \; .
\label{eq:comparaob}
\end{equation} 
The experimental determination of the slope $b$ of \kggs Eq.~(\ref{eq:bslope})
has been improved recently by the data on \kgee and \kgmm \cite{OA90,SP95}.
Both experiments use the BMS model in order to fit the slope from the data.
That is, they determine $\alpha_{K^*}$ defined by 
\begin{equation}
f(x) \, = \, \Frac{1}{1 -  x  \Frac{m_K^2}{m_{\rho}^2}}  +  
\Frac{\alpha_{K^*}}{1  -  x  \Frac{m_K^2}{m_{K^*}^2}} 
\, C_*  \left[  \Frac{4}{3}  -  
\Frac{1}{1  -  x  \Frac{m_K^2}{m_{\rho}^2}}  -  
\Frac{1}{9 (1  -  x  \Frac{m_K^2}{m_{\omega}^2})}  - 
\Frac{2}{9 (1  -  x  \Frac{m_K^2}{m_{\phi}^2})}  \right]
 \, , 
\label{eq:fxbms}
\end{equation}
where, as computed in \cite{BMS83},
\begin{equation}
C_* \; = \; \Frac{64}{3} \, \pi \, \alpha_{em} \, G_F \, m_V^2 \, f_V^3 \,
\Frac{h_V}{F_{\pi} \, |A_{\gamma \gamma}^{exp}|} \; \; \; \; ,
\label{eq:cstar}
\end{equation}
which numerical value is $C_* \simeq 3.1$ for $m_V = m_{\rho}$. 
However we will use the
numerical result given in \cite{OA90} that includes the $SU(3)$ breaking
$C_* = 2.5$. From the data the value of 
$\alpha_{K^*}$ in Eq.~(\ref{eq:fxbms}) can be extracted and it is 
\begin{eqnarray}
\alpha_{K^*} \, = \, - 0.28 \pm 0.13 \, & , & [K_L \rightarrow \gamma
e^+ e^-, \cite{OA90}] \nonumber \\
& & \label{eq:alphaks} \\
\alpha_{K^*} \, = \, - 0.03 \pm 0.11 \, & , & [K_L \rightarrow \gamma
\mu^+ \mu^-, \cite{SP95}] \; \; \; \; . \nonumber 
\end{eqnarray}
As can be seen the experimental situation still needs improvement.
We note the good agreement of the value of $\alpha_{K^*}$ from \kgee
with the prediction of the BMS model \cite{BMS83} $|\alpha_{K^*}| \simeq
0.2 \, - \, 0.3$.
\par
There is no reason a priori, according to us, that the form factor
in Eq.~(\ref{eq:fxbms}) should work better than just the linear 
expansion in Eq.~(\ref{eq:bslope}). Actually this last one is the one
suggested by $\chi$PT
\footnote{ See discussion in \cite{GE94}.}.
Of course Eq.~(\ref{eq:fxbms}) could be a 
resummation of higher order terms but it is not guaranteed to work. 
Deviations from the linear expansion are particularly important
for $K_L \rightarrow \gamma \mu^+ \mu^-$. Thus we suggest the 
experimentalists to fit $K_L \rightarrow \gamma e^+ e^-$ and 
$K_L \rightarrow \gamma \mu^+ \mu^-$ with both Eq.~(\ref{eq:bslope})
and Eq.~(\ref{eq:fxbms}) and then conclude which one works for 
both decays. Indeed if the linear slope is universal, then 
$\alpha_{K^*}$ determined from $K_L \rightarrow \gamma \mu^+ \mu^-$
should be substantially smaller than $\alpha_{K^*}$ measured
in $K_L \rightarrow \gamma e^+ e^-$, as suggested by the experimental
results in Eq.~(\ref{eq:alphaks}). As a first
approximation we determine
the full slope $b$, defined in 
Eq.~(\ref{eq:bslope}), 
experimentally from $K_L \rightarrow \gamma e^+ e^-$ by Taylor expanding
$f(x)$ in Eq.~(\ref{eq:fxbms}) and keeping only the linear slope. This
gives
\begin{equation}
b_{exp} \;  = \; 0.81 \pm 0.18 \; \; \; .
\label{eq:bexpsl}
\end{equation}
However one should emphasize that the coefficients of the quadratic
terms in Eq.~(\ref{eq:fxbms}) could be misleading since other 
contributions should be relevant at this chiral order
\footnote{We
have fitted a slope from the quadratic expansion of $f(x)$ in 
Eq.~(\ref{eq:fxbms}) and we find a $30 \%$
bigger value than the one quoted in Eq.~(\ref{eq:bexpsl}).}.
\par
Thus assuming $b_V^{nonet}$ in Eq.~(\ref{eq:rvst}) we can express the slope
$b_{D}$ Eq.~(\ref{eq:bsplit}) in terms of $\alpha_{K^*}$,
\begin{equation}
b_{D}^{exp} \, = \, - \, \Frac{4}{3} \, C_* \, \Frac{m_K^2}{m_V^2} \, 
\alpha_{K^*} \, \simeq \, 0.39 \pm 0.18  \; \; \; \; , 
\label{eq:bksal}
\end{equation}
where $m_{\phi}=m_{\omega}=m_{\rho}=m_V$ in 
Eqs.~(\ref{eq:fxbms}) has been used, and for the numerical
result we have put $m_V=m_{\rho}$ and $\alpha_{K^*} = -0.28 \pm 0.13$ 
as given by the \kgee data. We use this value since 
we think that the value of $\alpha_{K^*}$ from \kgee is 
definitely a better approach for the following reason: the value of 
$\alpha_{K^*}$ is fitted using the expression Eq.~(\ref{eq:fxbms}) that
has all orders in $x$ and a quick look to the expansion in $x$ shows that
the quadratic term is not so negligible against the slope. This is even
more important in the \kgmm case where the average $x$ is bigger than 
in \kgee.  
\par
The normalization specified in Eq.~(\ref{eq:aggs}) requires we take care
about our sign conventions. We use for the $K_L \rightarrow \gamma 
\gamma$ amplitude the same definition that in \kggs Eq.~(\ref{eq:klggs})
but now $A_{\gamma \gamma^*}(0) \, \equiv \, A_{\gamma \gamma}$. 
From the experimental width $K_L \rightarrow \gamma \gamma$ \cite{PDG94}
we get $|A_{\gamma \gamma}^{exp}| \, \simeq \, 
 3.45 \times 10^{-9} \, GeV^{-1}$. It is known that $K_L \rightarrow
\gamma \gamma$ is dominated by long distance contributions \cite{MAP81}.
By evaluating the 
diagram in Fig.~2.a  given by the pole model (\cite{DE95,MAP81,KLG84} and 
references 
therein) the leading $SU(3)$ \opc contribution vanishes due 
to the cancellation among $\pi^0$ and $\eta_8$. The experimental
magnitude of $A_{\gamma \gamma}$ shows that \ops contributions are 
very large. One could evaluate the diagram in Fig.~2.a using nonet 
symmetry with a mixing angle between $\eta_8$ and $\eta_1$,
$\theta_P \simeq - 20^{\circ}$ \cite{PDG94} and one would get 
a negative sign for $A_{\gamma \gamma}$.
As we shall see all the studied models require a positive value for
$A_{\gamma \gamma}$ in order to reproduce the experimental sign of
$b_D$ Eq.~(\ref{eq:bksal}). This fact
let us to think that a more thorough evaluation of the poorly known
$A(K_L \rightarrow \gamma \gamma)$ is needed and other \ops contributions
have to be taken into account, like a strong breaking of nonet
symmetry \footnote{There is no evidence of nonet symmetry breaking in 
$\eta' \rightarrow \gamma \gamma$ and $\eta' \rightarrow
\gamma \gamma^*$ according to Ref.~\cite{BB85} but 
see also \cite{HA96}.}.
Therefore since the pole model for $K_L \rightarrow \gamma \gamma$ does 
not seem to reproduce the right sign for the slope we input the sign
inferred from the experimental slope.

\subsection{Model predictions}
\hspace*{0.5cm}
As commented in the Introduction several models can be used in 
order to compute \ops vector meson exchange contributions to 
\kpiggtot and \kggs. We analyze critically their predictions for $a_V$
and $b_D$, most of which are calculated here for the first time.
\vspace*{0.5cm} \\
{\bf a) Factorization Model (FM). } 
\vspace*{0.4cm} \\
\hspace*{0.5cm}
The Factorization Model \cite{PI91,EK93,ENP94}
assumes that the dominant contribution of the four--quark operators
of the $\Delta S=1$ Hamiltonian comes from a factorization {\em current 
$\times$ current} of them. This assumption is implemented with a 
bosonization of the left--handed quark currents in $\chi$PT as given by
\begin{equation}
\overline{q_{jL}} \gamma^{\mu} q_{iL} \; \longleftrightarrow
\; L^{\mu} \, \equiv \, 
\Frac{\delta \, S [U,\ell,r,s,p]}{\delta \, \ell_{\mu, ji}} \; \; \; ,
\label{eq:boscur}
\end{equation}
where $i,j$ are flavour indices, $L_{\mu}$ is the chiral 
left--handed current and $S[U,\ell,r,s,p]$ is the 
low--energy strong effective action of QCD in terms of the Goldstone
bosons realization $U$ and the external fields $\ell, r, s, p$. 
\par
The general form of the lagrangian (assuming traceless left--handed
currents) is
\begin{equation}
{\cal L}_{FM} \; = \; 4 \, k_F \, G_8 \, \langle \, \lambda \, 
L_{\mu} \, L^{\mu} \, \rangle \; + \; h.c. \; \; \; , 
\label{eq:fmgenera}
\end{equation}
where $\lambda \, \equiv \, \frac{1}{2} \, (\lambda_6 \, - \, i \lambda_7)$
and
$
L_{\mu} \, = \, L_{\mu}^1 \, + \, L_{\mu}^3 \, + \, L_{\mu}^5 \, + ...
$
which first term $L_{\mu}^1$ was already 
given in Eq.~({\ref{eq:leftcu}).
The FM  gives a full prediction but for a fudge factor
$k_F$ that is not given by the model and in general $k_F \sim 
{\cal O}(1)$. The naive factorization
approach (nFM)  
($k_F = 1$) would correspond to inputting the $\Delta I = 1/2$ 
enhancement once the left--handed currents are generated by the full
relevant strong lagrangian. Thus caution is needed in order to
interpret this factor.
\par
For \kpiggtot the FM lagrangian
density is 
\begin{equation}
{\cal L}_{FM} \, = \, 4 \, k_F \, G_8 \, \langle \, \lambda \, \left\{ \, 
\Frac{\delta S_2^{\chi}}{\delta \ell^{\mu}} \, , \, 
\Frac{\delta S_V^{PP\gamma \gamma}}{\delta \ell_{\mu}} \, \right\} \rangle
\; + \; h.c. \; \; , 
\label{eq:lfm}
\end{equation}
where $S_2^{\chi}$ is the action associated to the leading ${\cal O}(p^2)$
strong $\chi$PT lagrangian Eq.~(\ref{eq:str2}) and $S_V^{PP\gamma\gamma}$ 
is the action of the strong 
\ops lagrangian obtained by integrating out the vector mesons between
two vertices generated by ${\cal L}(VP\gamma)$ in Eq.~(\ref{eq:vpga}). 
In Eq.~(\ref{eq:lfm}) the left--handed current associated to 
$S_V^{PP\gamma\gamma}$ is not automatically traceless. Since we are 
considering only 
octet currents in our evaluation we have subtracted the trace term
\footnote{We thank G. Ecker for calling our attention about this point.}.
\par
Then we find
\begin{eqnarray}
a_{V,0}^{dir} \,  =  \, - \, 4 \, k_F \, a_{V,0}^{ext} 
\; \; \; \; \; \; \; & , &
\; \; \; \; \; \; \; 
a_{V,+}^{dir} \,  =  \, - \, 2 \, k_F \, a_{V,+}^{ext} \; \;   .
\label{eq:avfm}
\end{eqnarray}
As $k_F >0$, the FM predicts a cancellation between
both contributions to $a_V$. The results are collected in Table I.
\par
In order to evaluate the slope in the factorization model we first
construct the strong lagrangian generated by vector exchange between
${\cal L}(VP\gamma)$ Eq.~(\ref{eq:vpga}) and ${\cal L}(V\gamma)$ 
Eq.~(\ref{eq:fvvg}). This gives
\begin{equation}
{\cal L}(P\gamma \gamma^*) \, = \, - \, 
\Frac{h_V \, f_V}{\sqrt{2} \, m_V^2} \, 
\varepsilon_{\mu \nu \rho \sigma} \, \langle \, 
\nabla_{\alpha} \, f_{+}^{\alpha \mu} \, \{ \, u^{\nu} \, , \, 
f_{+}^{\rho \sigma} \, \} \, \rangle \; \; \; \; . 
\label{eq:vklggs}
\end{equation}
The FM lagrangian density for \kggs can be read from the relevant
pieces in Eq.~(\ref{eq:fmgenera}).
We get for the octet and nonet case respectively,
\begin{eqnarray}
b^{octet}_{D} \, &  = & \, \Frac{256 \sqrt{2}}{9} \, \pi \, G_8 \, F_{\pi} \, 
h_V \, f_V \, \alpha_{em} \, \Frac{m_K^2}{m_V^2} \, 
\Frac{1}{|A_{\gamma \gamma}^{exp}|} \, k_F \; \; \; \; , \nonumber \\
& & \label{eq:bkstfm} \\
b^{nonet}_{D} \, & = & \, 2 \, b^{octet}_{D} \; \; \; \; . \nonumber 
\end{eqnarray}
We note that as $h_V \, f_V \, > \, 0$, $b_{D} > 0$ for $k_F > 0$.
Moreover, for $0 \, < \, k_F \, < \, 1$ we find $0 \, < \, b^{octet}_{D} \, 
< \, 0.71$, $0 \, < \, b^{nonet}_{D} \, 
< \, 1.42$. The upper limit corresponds to nFM ($k_F = 1$). 
The results for the slope are
collected in Table II.
We could compare our expressions for 
$b_D$ in Eq.~(\ref{eq:bkstfm}) with $b_D^{exp}$ in 
Eq.~(\ref{eq:bksal}). However the determination of $b_D^{exp}$ 
in Eq.~(\ref{eq:bksal}) is very much model dependent.
In order to minimize this dependence we are forced to compare with 
$b_{exp} - b_V$ in Eqs.~(\ref{eq:rvst},\ref{eq:bocte},\ref{eq:bexpsl})
instead. We get
\begin{eqnarray}
b_{exp} \, - \, b_V^{octet} \, & = & \, b_D^{octet} \; \; \; \longrightarrow
\; \; \; k_F \, = \, 1.15 \pm 0.25 \; \; \; , \nonumber \\
b_{exp} \, - \, b_V^{nonet} \, & = & \, b_D^{nonet} \; \; \longrightarrow
\; \; \; k_F \, = \, 0.25 \pm 0.13 \; \; \; .  \label{eq:kfkstar} 
\end{eqnarray}
Indeed the $FM$ nonet solution in Eq.~(\ref{eq:kfkstar}) gives a very
small value of $k_F$ unacceptable for \kpiggn. The first solution 
(octet) seems more reasonable from a theoretical point of view.
Analogous conclusions would have been obtained by fixing $k_F$ from
$a_{V,0}$.
\vspace*{0.5cm} \\
{\bf b) Bergstr\"om--Mass\'o--Singer (BMS) model. }
\vspace*{0.4cm} \\
\hspace*{0.5cm}
The basic idea of the BMS model \cite{BMS83} is to suggest that the
structure of the weak $VP\gamma$ vertex is dominated by a weak 
vector--vector transition as shown in Fig.~1.e. That is, for example, 
the $K \rho \gamma$ coupling
would be generated through the sequence $K \rightarrow K^* \gamma$, 
$K^* \rightarrow \rho$. 
This model was introduced in \cite{BMS83} in the study of the 
\kggs form factor. It is argued that the weak vector--vector transition
lacks penguin contributions because of the scalar and pseudoscalar
quantum numbers of the penguin operators \footnote{The penguin 
$(V-A) \times (V+A)$ structure can be Fierz ordered to $(SS-PP)$.}.
Following this approach we introduce the leading effective
weak vector--vector transition as 
\begin{equation}
{\cal L}_W(VV) \, = \, 4 \, \beta \, G_8 \, F_{\pi}^2 \, m_V^2 \, 
\langle \, \Delta  \, V_{\mu} V^{\mu} \, \rangle \; \; \; \; ,
\label{eq:wvv}
\end{equation}
where $\beta$ is an, in principle,  unknown
dimensionless coupling constant that fixes the strength of the 
transition.
\par
Therefore we can evaluate the amplitudes for \kpiggtot using 
${\cal L}(VP\gamma)$ Eq.~(\ref{eq:vpga}) and give the following 
predictions for $a_V$~:
\begin{eqnarray}
a_{V,0}^{dir} \,  =  \, - \, \beta \, a_{V,0}^{ext} 
\; \; \; \; \; \; \; & , & \; \; \; \; \; \; \; 
a_{V,+}^{dir} \, =  \, \, \beta \, a_{V,+}^{ext} \; \;  .
\label{eq:avkap}
\end{eqnarray}
Then with ${\cal L}(VP\gamma)$ Eq.~(\ref{eq:vpga}) and 
${\cal L}(V\gamma)$ Eq.~(\ref{eq:fvvg}) we can evaluate the slope in the
BMS model. It reads
\begin{equation}
b_{D} \, = \, \Frac{512 \sqrt{2}}{9} \, \pi \, G_8 \, F_{\pi} \, 
h_V \, f_V \, \alpha_{em} \Frac{m_K^2}{m_V^2} \, 
\Frac{1}{|A_{\gamma \gamma}^{exp}|} \, \beta \; \; \; \; . 
\label{eq:bksbms}
\end{equation}
Motivated by the paper of Shifman, Vainshtein and Zakharov 
\cite{VZS77}, where the penguin diagram was advocated to explain the 
$K \rightarrow \pi \pi$ matrix elements,  another
basic feature of the BMS model is to assume
that the weak vector--vector transition has no $\Delta I = 1/2$
enhancement and hence $\beta$ can be reliably evaluated
using factorization and vacuum insertion. 
Bergstr\"om, Mass\'o and Singer \cite{BMS83} got
$| \beta | \, \simeq \, 0.2 \, - \, 0.3$. In fact by comparing
Eq.~(\ref{eq:bksbms}) with Eq.~(\ref{eq:bksal}) we obtain 
\begin{equation}
\beta \, \simeq \, - \, \alpha_{K^*} \, = \,  0.28 \pm 0.13 \; \; \; \;
\; .
\label{eq:kappaalp}
\end{equation}
From the experimental value of $\alpha_{K^*}$ we see that the magnitude
of the coupling $\beta$ predicted in \cite{BMS83} is in good
agreement with the experiment on the channel \kgee.
We get the prediction of the model for 
$a_V$ collected in Table I. We conclude then that the BMS model
is not able to give a consistent picture of both processes.
\vspace*{0.5cm} \\
{\bf c) Weak Deformation Model (WDM). }
\vspace*{0.4cm} \\
\hspace*{0.5cm} 
The WDM \cite{EP90,ECG90} assumes that the dominant effect at long
distances of the left--handed $\Delta S=1$ Lagrangian can be obtained
through a deformation of the two 1--forms defined in the coset space
of the spontaneously broken chiral group. 
Even when
the dynamics behind the WDM is not clear its phenomenological applications
might be successful and in any case they deserve to be tested.
\par
It has been shown in \cite{EK93} that the WDM can be expressed through
the lagrangian
\begin{equation}
{\cal L}_{WDM} \; = \; 2 \, G_8 \, \langle \, \lambda \,
\left\{ \, L_{\mu}^1 \, , \, L^{\mu} \, \right\}
\, \rangle \; + \; h.c. \; \; \; , 
\label{eq:wdmgen}
\end{equation}
and comparing with ${\cal L}_{FM}$ in Eq.~(\ref{eq:fmgenera}) one sees
that, at \opc, ${\cal L}_{WDM} \, = \, {\cal L}_{FM} (k_F=1/2)$, but at
higher chiral orders the FM can have additional terms.
This equivalence can be 
extended at \ops \footnote{G. Ecker, private communication.} if
the lagrangian is still a product of currents ${\cal O}(p)$ and 
${\cal O}(p^5)$.
\par
The WDM gives a definite prediction for $a_V^{dir}$. This 
is (for an octet current) \cite{EP90}
\begin{eqnarray}
a_{V,0}^{dir} \,  =  \, - \, 2 \, a_{V,0}^{ext} 
\; \; \; \; \;  \; \; & , & \; \; \; \; \; \; \; 
a_{V,+}^{dir} \,  =  \, -  \, a_{V,+}^{ext} \; \;  ,
\label{eq:avdwdm}
\end{eqnarray}
showing a strong cancellation that is complete for \kpiggp. 
\par
The procedure necessary to get the slope in the WDM consists in 
\lq\lq deformating" the octet current
coupled to vector mesons in ${\cal L}(VP\gamma)$ Eq.~(\ref{eq:vpga})
and to integrate out the vector meson interchanged with a vertex
generated by ${\cal L}(V \gamma)$ Eq.~(\ref{eq:fvvg}). We get from 
diagram in Fig.~2.c
\begin{equation}
b^{octet}_{D} \, = \, \Frac{128 \sqrt{2}}{9} \, \pi \, G_8 \, F_{\pi}
\, h_V \, f_V \, \alpha_{em} \Frac{m_K^2}{m_V^2} \, 
\Frac{1}{|A_{\gamma \gamma}^{exp}|} \, \simeq \, 0.35 \; \; \; \; .
\label{eq:bwdm}
\end{equation}
This result has also been obtained by Ecker \cite{ECK90} who added it
to $b_V^{nonet}$ in Eq.~(\ref{eq:rvst}). However we think that 
$b_D^{octet}$ should be added up to $b^{octet}_V = 0$, and therefore
the experimental result Eq.~(\ref{eq:bexpsl}) is not recovered. 
The extension 
to the nonet case is ambiguous for this model \cite{EK93}, though the
consistent choice, which enlarges the equivalence theorem between \opc
FM with $k_F = 1/2$ and WDM, to the nonet, would lead to
$
b^{nonet}_D \, \simeq \, 0.70 \, .
$
This result added to $b^{nonet}_V$ in Eq.~(\ref{eq:rvst}) could still
not to be excluded. In any case this is not a firm prediction of the
WDM.

\section{Factorizable Wess--Zumino--Witten anomaly contribution from
vector resonances to \kpiggtot and \kggs}
\hspace{0.5cm}
The procedure used in the FM to study the resonance exchange contribution
to a specified process involves first the construction of a 
resonance exchange generated strong Lagrangian, in terms of Goldstone bosons
and external fields, from which to evaluate the left--handed currents.
For example, in \kpiggtot one starts from the strong $VP\gamma$ vertex
and integrates out the vector mesons between two of those vertices 
giving a strong Lagrangian for $P P \gamma\gamma$ ($P$ is short for
pseudoscalar meson) from which to evaluate the left--handed current.
This method of implementing the FM imposes as a constraint
the strong effective $P P \gamma \gamma$ vertex and therefore
it can overlook parts of the chiral structure of the weak vertex.
In the FM model the dynamics is hidden in the fudge factor $k_F$. 
Alternatively one could try to identify the different vector meson
exchange contributions and then estimate the relative weak
couplings.
\par
Instead of getting the strong lagrangian generated
by vector meson exchange we apply the factorization procedure to the
construction of the weak $VP\gamma$ vertex and we integrate out the
vector mesons afterwards. The interesting advantage of our approach is that, 
as we will see in the processes we are interested in, allows us to 
identify new contributions to the left--handed currents and therefore
to the chiral structure of the weak amplitudes.
\par
Let us specify the procedure.
For later use we will split the strong effective action $S$ and 
the corresponding left--handed current ${\cal J}_{\mu}$ in two
pieces~: $S = S_1 + S_2$ and ${\cal J}_{\mu} = {\cal J}^1_{\mu} + 
{\cal J}^2_{\mu}$, respectively.
The bosonization of the four--quark operators in 
${\cal H}_{NL}^{\Delta S=1}$ proposed in \cite{PI91} and specified
in Eq.~(\ref{eq:boscur}) tells us
\begin{equation}
\overline{q_l}\gamma^{\mu}(1-\gamma_5)q_k \overline{q_j} \gamma_{\mu}
(1-\gamma_5) q_i \, \longleftrightarrow \, 
4 \, [ ({\cal J}_{\mu}^1)_{lk} \, ({\cal J}_2^{\mu})_{ji} \, + \, 
       ({\cal J}_{\mu}^2)_{lk} \, ({\cal J}_1^{\mu})_{ji} \, ]
\; \; \; , 
\label{eq:bosoniza}
\end{equation}
with $i,j,k,l$ flavour indices (terms ${\cal J}_{\mu}^1 {\cal J}^{\mu}_1$
or ${\cal J}_{\mu}^2 {\cal J}^{\mu}_2$ have been dropped since they 
will not contribute to the processes we are considering as we shall see later).
It can be shown that, in this factorizable approach, the $Q_{-}$ operator
defined in Eq.~(\ref{eq:qplusm}) is represented by 
\begin{equation}
Q_{-} \, \longleftrightarrow \, 4 \, \left[ \, \langle \lambda
\, \{ {\cal J}_{\mu}^1 , {\cal J}^{\mu}_2 \, \} \rangle \, - \, 
\langle \lambda {\cal J}_{\mu}^1 \rangle \langle {\cal J}^{\mu}_2 \rangle \, 
- \, \langle \lambda {\cal J}_{\mu}^2 \rangle \langle {\cal J}^{\mu}_1 
\rangle \, \right] \; \; \; , 
\label{eq:qmresult}
\end{equation}
where, for generality, the currents have been supposed to have
non--zero trace \footnote{A similar procedure in the study of the chiral
anomaly in non--leptonic weak interactions was used in \cite{BE92}.}.
\par
If we want to apply this procedure in order to construct the
factorizable contribution to the ${\cal O}(p^3)$ weak $VP\gamma$ 
vertex we have to identify in the full strong action which pieces
(at this chiral order) can contribute. Moreover for \kpiggtot and 
\kggs we are interested in a weak $VP\gamma$ vertex with the 
Levi--Civita antisymmetric pseudotensor. 
\par
According to our study it turns out that there are four terms in 
the strong action to take into account and which Lagrangian density is 
given in Eqs.~(\ref{eq:str2},\ref{eq:zulr},\ref{eq:vpga} and 
\ref{eq:fvvg}). Analogously to the specified procedure we define 
correspondingly
\begin{eqnarray}
S & = & \, S_V \, + \, S_{\varepsilon} \; \; \; \; , \nonumber \\
S_V & = &  \, S(V\gamma) \, + \, S_2^{\chi}  \; \; \; \; , 
\label{eq:actions} \\
S_{\varepsilon} & = & S_{WZW} \, + \, S(VP\gamma) \; \; \; \; , \nonumber
\end{eqnarray}
where the notation is self-explicative. Constructing the left--handed
currents and keeping only the terms of interest we get
\begin{eqnarray}
\Frac{\delta \, S_V}{\delta \ell_{\mu}} \, & \; = \; & 
\, - \, \Frac{f_V}{\sqrt{2}} m_V^2 \, u^{\dagger} \, V^{\mu} \, u \, - \, 
\Frac{1}{2} F_{\pi}^2 u^{\dagger} \, u^{\mu} \, u \; \; \; , \nonumber \\
& & \label{eq:currveps} \\
\Frac{\delta \, S_{\varepsilon}}{\delta \ell_{\mu}} \, & \; = \; & 
\, \varepsilon^{\mu \nu \alpha \beta} \, \left[ \, \Frac{1}{16 \pi^2}
\left\{ F_{\nu \alpha}^L + \Frac{1}{2} U^{\dagger} F_{\nu \alpha}^R
U \, , \, u^{\dagger}  \, u_{\beta} \, u \, \right\} \, + \, 
h_V \, u^{\dagger} \left\{ f^{+}_{\nu \alpha} \, , \, V_{\beta} \right\} u
\, \right] \; \; \; .
\nonumber
\end{eqnarray}
Note that if we are assuming nonet symmetry in the vector meson sector
the first term in $\delta S_V / \delta \ell_{\mu}$ is not traceless. 
The first term in $\delta S_{\varepsilon} / \delta \ell_{\mu}$ obtained
from the Wess--Zumino--Witten action Eq.~(\ref{eq:zulr}) is not 
traceless either, (because the anomalous breaking of the chiral symmetry),
while the last term in $\delta S_{\varepsilon} / \delta \ell_{\mu}$ has
to be taken traceless or not according to the inclusion of octet or nonet
currents respectively. We will discuss these cases later on. From 
Eq.~(\ref{eq:currveps}) we see that when the first term of the first current
couples with the first term of the second current and the second terms
correspondingly, we get two different contributions to the weak $VP\gamma$
vertex.
\par
By applying the factorization procedure Eq.~(\ref{eq:qmresult}) to our
currents Eq.~(\ref{eq:currveps}) we get our effective action
\begin{eqnarray}
{\cal L}_W^{fact}(VP\gamma) \, & = & \, 4 \, G_8^{eff} \, 
\left[ \, \langle \, \lambda \, \left\{ \Frac{\delta S_V}{\delta \ell^{\mu}} \,
, \, \Frac{\delta S_{\varepsilon}}{\delta \ell_{\mu}} \, \right\} \rangle
\, - \, \langle \lambda \Frac{\delta S_V}{\delta \ell^{\mu}} \rangle
\langle \Frac{\delta S_{\varepsilon}}{\delta \ell_{\mu}} \rangle \right.
\nonumber \\
& & \; \; \; \; \; \; \; \; \; \; \; \; \; \left.
\, - \, \langle \lambda \Frac{\delta S_{\varepsilon}}{\delta \ell^{\mu}} 
\rangle \langle \Frac{\delta S_V}{\delta \ell_{\mu}} \rangle \, \right] 
\; + \, h.c. \; \; \; \; . 
\label{eq:anomvpg} 
\end{eqnarray}
In Eq.~(\ref{eq:anomvpg}) we have defined the effective coupling 
$G_8^{eff}$, since it does not necessarily coincide with $G_8$ in 
Eq.~(\ref{eq:g8fr}). This is a free coupling in our model. For definiteness
and, as we shall see, consistently with phenomenology, we use for 
$G_8^{eff}$ the naive value that would arise from the 
$C_{-}(m_{\rho})$ Wilson coefficient (with ${\cal O}(\alpha_s)$ 
corrections included) in Eq.~(\ref{eq:g81}), i.e.  
\begin{eqnarray}
G_8^{eff} \,   =  \, \eta \, G_8 
\; \; \; \; \; \; & , & \; \; \; \; \; \; 
\eta \,  =  \Frac{g_8^{Wilson}}{|g_8|_{K \rightarrow \pi \pi}} 
\; \simeq \; 0.21 \; \; \; , 
\label{eq:eta8}
\end{eqnarray}
with $|g_8|_{K \rightarrow \pi \pi}$ and $g_8^{Wilson}$ defined 
in Eqs.~(\ref{eq:g8fr},\ref{eq:g81})
respectively. That is we do not input any $\Delta I = 1/2$ enhancement
in our model. However there is no reason {\em a priori} for this choice
(but neither to exclude) and one could keep this as a free parameter
analogously to $k_F$ in the FM model. We call our model
prescribed by Eqs.~(\ref{eq:anomvpg}, \ref{eq:eta8}) the Factorization
Model in the Vector couplings (FMV).
\par
By inputting Eq.~(\ref{eq:currveps}) into Eq.~(\ref{eq:anomvpg}), 
expanding and comparing with ${\cal L}_W(VP\gamma)$ in Eq.~(\ref{eq:mostg})
we can determine the contributions of our FMV model to the couplings
$\kappa_i, i=1,2,3,4,5$ defined in Eq.~(\ref{eq:fullcur}). In order
to simplify our expressions we define
\begin{equation}
\ell_V \, \equiv \, \Frac{3}{16 \sqrt{2} \pi^2} \, f_V \, 
\Frac{m_V^2}{F_{\pi}^2} \, \; \; \; . 
\label{eq:rhov}
\end{equation}
Then, assuming only octet currents (i.e. the term in $h_V$ in the 
second current of Eq.~(\ref{eq:currveps}) is made traceless) we get
\begin{eqnarray}
\kappa_1^{FMV} \, & = & \, 0 \; \; \; \; \; , \nonumber \\
\kappa_2^{FMV} \, & = & \, ( \, 2 \, h_V \, - \, \ell_V \, ) \,  \eta \; \; 
\; \; \; , 
\nonumber \\
\kappa_3^{FMV} \, & = & \, - \, \Frac{8}{3} \, h_V \, \eta \; \; \; \; 
\; , \label{eq:kappanom1} \\
\kappa_4^{FMV} \, & = & \,  2 \, \ell_V \, \eta \; \; \; \; \; ,
\nonumber \\ 
\kappa_5^{FMV} \, & = & \,  2 \, \ell_V \, \eta \; \; \; \; \; ,
\nonumber  
\end{eqnarray}
while if we assume nonet currents only $\kappa_3$ changes and it 
becomes
\begin{equation}
\kappa_3^{FMV} \, |_{nonet} \; = \; - \, 4 \, h_V \, \eta \; \; \; . 
\label{eq:ka3n}
\end{equation}
With these results we can give now a prediction from the FMV model to 
both $a_V^{dir}$ Eq.~(\ref{eq:avdirk}) and the slope $b_D$
in Eq.~(\ref{eq:bstar}),
\begin{eqnarray}
a_{V,0}^{dir} \, \left|_{FMV} \, \right. \; & = & \; 
- \, \Frac{1024 \pi^2}{9} \, h_V \, \Frac{m_K^2}{m_V^2} \, 
\left[ \, 2 h_V \, + \, \ell_V \, \right] \, \eta \; \simeq \; -0.95 \; \; ,
\nonumber \\
& & \label{eq:avdirre} \\
a_{V,+}^{dir} \, \left|_{FMV} \, \right. \; & = & \; 
- \, \Frac{128 \pi^2}{9} \, h_V \, \Frac{m_K^2}{m_V^2} \, 
\left[ \, \ell_V \, - \, 2 h_V \, \right] \, \eta \; \simeq \; -0.05 \; \; , 
\nonumber 
\end{eqnarray}
and
\begin{eqnarray}
b_D^{octet} \; \left|_{FMV} \, \right. \; & = & \; 
 \, \Frac{128 \sqrt{2}}{9} \, \pi \, G_8 \, \alpha_{em} \, F_{\pi}
 \, f_V \, \Frac{m_K^2}{m_V^2} \, \Frac{1}{|A_{\gamma \gamma}^{exp}|}
\left[ \, 2 h_V \, + \, \ell_V \, \right] \, \eta \; \simeq \; 0.51 \; \; ,
\nonumber \\
& & \label{eq:bstaranom} \\
b_D^{nonet} \; \left|_{FMV} \, \right. \; & = & \; 
 \, \Frac{128 \sqrt{2}}{9} \, \pi \, G_8 \, \alpha_{em} \, F_{\pi}
 \, f_V \, \Frac{m_K^2}{m_V^2} \, \Frac{1}{|A_{\gamma \gamma}^{exp}|}
\left[ \, 4 h_V \, + \, \ell_V \, \right] \, \eta \; \simeq \; 0.66 \; \; .
\nonumber 
\end{eqnarray}
Before proceeding on, it is worth to remark the similarities and 
differences between our FMV and the usual FM approach. We have already
stated above that the basic difference in our model is that we 
apply the FM to construct a weak $VP\gamma$ vertex and then we 
integrate out the vector mesons afterwards. 
If we take a look to $a_V^{dir} (FMV)$ in Eq.~(\ref{eq:avdirre}) and 
compare it with the FM result Eq.~(\ref{eq:avfm}) or  
the WDM result Eq.~(\ref{eq:avdwdm}) we see that the part 
from $a_V^{dir} (FMV)$
in $h_V^2$ coincides with the FM and the WDM for
$\eta = 1/2$, that is $G_8^{eff} = G_8 / 2$ or $k_F = 1/2$ in the FM.
Therefore the term in $\ell_V$ in $a_V^{dir}(FMV)$ Eq.~(\ref{eq:avdirre})
is a complete {\bf new} contribution. Exactly the same applies to 
$b_D^{octet} (FMV)$.
\par
The relevance of our new contribution goes in fact further than a 
numerical result. Let us consider the effective lagrangian for 
\kpiggtot generated by vector resonance exchange between a vertex
from ${\cal L}_W^{fact}(VP\gamma)$ Eq.~(\ref{eq:anomvpg}) and a 
vertex from ${\cal L}(VP\gamma)$ in Eq.~(\ref{eq:vpga}). We arrive to
\begin{eqnarray}
{\cal L}_{eff}^{K \rightarrow \pi \gamma \gamma} \; & = & \; 
4 \pi \alpha_{em} \, h_V \, G_8^{eff} \, \Frac{F_{\pi}^2}{m_V^2} \, 
\left[ \, \varepsilon^{\mu \nu \alpha \beta} \, 
\varepsilon_{\mu \gamma \delta \epsilon} \, \right]  \cdot
\nonumber \\
& & \; \left[ \, ( \, \ell_V \, - \, 2 h_V \, ) \, 
\langle \, \{ u_{\nu} , f^+_{\alpha \beta} \} \, \{ \, \{ \Delta , 
\,  u^{\gamma} \} \,  , \,  f_+^{\delta \epsilon} \, \} \, \rangle \right.
\nonumber \\
& & \; \left. \; \; + \, \Frac{8}{3} \, h_V \, \langle \, \Delta
\, u^{\gamma} \, \rangle \, \langle \, f_+^{\delta \epsilon} \, 
\{ \, u_{\nu} \, , \, f^+_{\alpha \beta} \, \} \, \rangle \; \right.
\label{eq:leffkpi} \\
& & \; \left. \; \; - \, 4 \, \ell_V \, \langle \, \Delta \, 
\{ \, u_{\nu} \, , \, f^+_{\alpha \beta} \, \} \, \rangle \, 
\langle \, f_+^{\delta \epsilon} \, u^{\gamma} \, \rangle \, \right] 
\; \; \; \; \; . 
\nonumber 
\end{eqnarray}
As can be seen the last term in Eq.~(\ref{eq:leffkpi}) is only proportional
to $\ell_V$ and therefore it gives a new chiral structure operator to 
the process that has not been taken into account before.
This is particularly evident from formulae in Eq.~(\ref{eq:avdirre})
for $a_{V,0}$ and $a_{V,+}$ where different Clebsch--Gordan relations
weight the terms proportional to $\ell_V$ and $h_V$.
This model proves to be more efficient that the usual FM approach in that
it is able to identify a new chiral structure that contributes to both
\kpiggtot and \kggs. When this piece is included we find that a consistent
picture of these channels arises. In particular, we find agreement between
the phenomenologically expected \ops vector meson contribution to \kpiggn
and our prediction.
\par
It is interesting to note that our FMV model result in Eq.~(\ref{eq:currveps})
can also be derived in the Hidden symmetry formulation of vector mesons
\cite{BK88} (see the Appendix). As already noticed previously \cite{BB85}
the phenomenological value of $h_V$ can be nicely reproduced by the so 
called \lq\lq complete vector meson dominance". As we show in the Appendix, 
in this scheme one obtains also the relation
$
\ell_V \, = \, 4  h_V \, , 
$
phenomenologically consistent. Then our results can be easily derived. 
Thus, from this point of view, the Hidden symmetry formulation with
\lq\lq complete vector meson dominance" seems a complete and predictive
framework.

\begin{sidewaystable}
\begin{center}
\begin{tabular}{|c|c|c|c|c|c|c|c|c|}
\multicolumn{9}{c}{{\bf Table I}} \\
\multicolumn{9}{c}{} \\
\hline
\hline
\multicolumn{1}{|c|}{} &
\multicolumn{1}{|c|}{} &
\multicolumn{4}{|c|}{$a_V^{dir}$} &
\multicolumn{3}{|c|}{$a_V = a_V^{ext} + a_V^{dir}$} \\ \cline{3-9}
Channel & $a_V^{ext}$ & & &  & & & &  \\
\multicolumn{1}{|c|}{} &
\multicolumn{1}{|c|}{} &
\multicolumn{1}{|c|}{WDM} &
\multicolumn{1}{|c|}{FM} &
\multicolumn{1}{|c|}{BMS} &
\multicolumn{1}{|c|}{FMV} &
\multicolumn{1}{|c|}{WDM} &
\multicolumn{1}{|c|}{FM} &
\multicolumn{1}{|c|}{FMV + BMS} \\
\hline
\hline
& & & & & & & &  \\
\kpiggn & 0.32  & $-$0.64 & $-$1.28 $k_F$   & $-$0.09 & $-$0.95 & $-$0.32 & 
0.32 $-$ 1.28 $k_F$ & $-$0.72  \\
& & & & & & & &  \\
\hline
& & & & & & & &  \\
\kpiggp & $-$0.08 & 0.08 & 0.16 $k_F$ & $-$0.02 & $-$0.05 & 0 & 0.16 $k_F$
$-$ 0.08 & $-$0.15  \\
& & & & & & & &  \\
\hline
\hline
\end{tabular} 
\caption{Results for $a_V^{dir}$ from WDM, FM, BMS and FMV models
and final result for the $a_V$ parameter in both \kpiggn and \kpiggp 
channels. The result of the FMV model is for $\eta \, = \, 0.21$ 
as given in Eq.~(\ref{eq:eta8}). For the FM results the values of
$k_F$ would be the ones in Eq.~(\ref{eq:kfkstar}) for the different
schemes~: octet or nonet.}
\end{center}
\end{sidewaystable}

\section{Results}
\hspace*{0.5cm}
In this Section we are going to collect and discuss our results. 
In our approach we add the different vector meson exchange contributions
to the processes under consideration. We have
already seen that the BMS model by itself only can account for the
slope $b_D$ in \kggs but not for $a_{V,0}$ in \kpiggn. In fact if we 
compare diagrams in Fig.~1.c (with a direct weak $VP\gamma$ vertex) 
and in Fig.~1.e (with the weak V--V transition, BMS model) (or Fig.~2.c 
with Fig.~2.d) we realize that both contributions are
independent because they have different physical mechanism and 
analytical structure (the BMS
amplitude has an extra pole due to the two vector propagators).
This is the reason why we conclude that both
contributions, the BMS and FMV, are independent and therefore should
be added up. This addition is numerically
relevant for \kggs but not in \kpiggtot.

\subsection{\kpiggtot}
\hspace*{0.5cm} 
Our results for $a_V$ have been written in Table I. As a final 
result we have
\begin{eqnarray}
a_{V,0} \, & = & \, a_{V,0}^{ext} \, + \, a_{V,0}^{dir} \, |_{FMV} \, + \,
a_{V,0}^{dir} \, |_{BMS} \; \simeq \; -0.72 \; \; \; , \nonumber \\
& & \label{eq:avtotanom} \\
a_{V,+} \, & = & \, a_{V,+}^{ext} \, + \, a_{V,+}^{dir} \, |_{FMV} \, + \,
a_{V,+}^{dir} \, |_{BMS} \; \simeq \; -0.15 \; \; \; . \nonumber 
\end{eqnarray}
As shown
in \cite{CE93}, once unitarity corrections are included \cite{CD93,CE93},
a value of $a_{V,0} \simeq -0.9$ is able to reproduce well the width and
spectrum of \kpiggn. 
Lately Holstein and Kambor \cite{KH94} have studied a Khuri--Treiman 
unitarization of the \kpiggn through the
$K \rightarrow \pi \pi \pi$ intermediate states
\footnote{In \cite{KH94} only the dominant S--wave component in 
$\pi \pi$ scattering has been taken into account and therefore only 
and A--type amplitude is generated by these corrections.},
and the inclusion of the
experimental $\gamma \gamma \rightarrow \pi^0 \pi^0$.
This modifies 
slightly this value to about $a_{V,0} \simeq  -0.8$.

\begin{figure}
\begin{center}
\leavevmode
\hbox{%
\epsfxsize=14cm
\epsffile{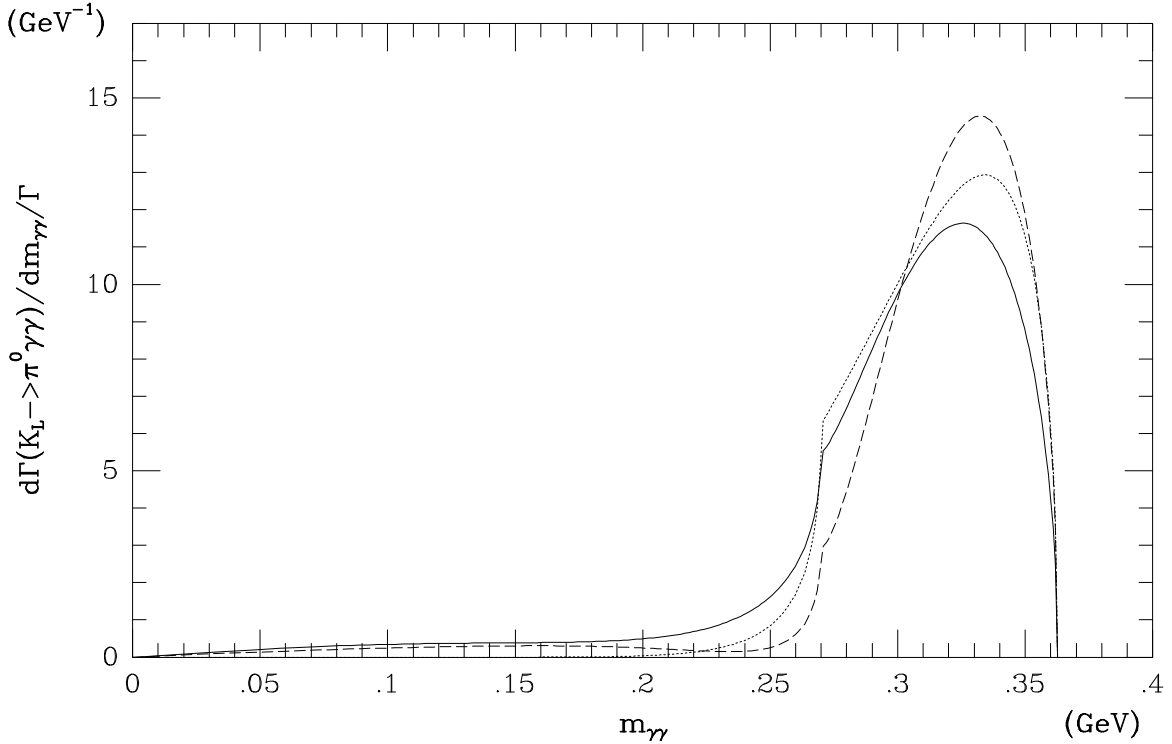}}
\end{center}
\vspace*{-10cm}
\caption{Normalized diphoton invariant mass spectrum for \kpiggn
at \opc (dotted line), \ops with $a_{V,0}=0$ (dashed line) and 
\ops with $a_{V,0}=-0.7$ (full line). The \ops curves also include the
unitarity corrections from $K_L \rightarrow \pi^{0} \pi^+ \pi^-$
from \protect\cite{CD93,CE93} and the inclusion of experimental 
$\gamma \gamma \rightarrow \pi^{0} \pi^{0}$ 
\protect\cite{KH94}.}
\end{figure}

For the central value of our result Eq.~(\ref{eq:avtotanom}), 
$a_{V,0} \simeq -0.7$ we find
\begin{equation}
Br( \kpiggne ) \; = \; 1.50 \, \times \, 10^{-6} \; \; \; \; , 
\label{eq:brkltheofin}
\end{equation}
to be compared with the experimental world average $Br (\kpiggne)_{exp} 
\, = \, (1.70 \pm 0.28) \, \times \, 10^{-6}$ \cite{PDG94} (see also
Eq.~(\ref{eq:brklexp})). In Fig.~3 we compare the normalized diphoton
invariant mass spectrum of \kpiggn at \opc and \ops with 
$a_{V,0} \, = \, 0, -0.7$.
\par
As already emphasized earlier, this prediction is based on taking for
the parameter $\eta$ Eq.~(\ref{eq:eta8}) the value predicted by the 
Wilson coefficient in Eq.~(\ref{eq:g81}). Though it is a very 
interesting and predictive feature of the model and supported by
experiment, it is not guaranteed to work. However one could fix the
$\eta$ parameter by the experimental slope of \kggs and then one 
would find still the same or, may be, a slightly larger value.
For a complete understanding of the underlying quark dynamics
one should enlarge our model to other channels and then find
the effective coupling analogous to the one in Eq.~(\ref{eq:eta8}).
\par
In Fig.~4 we plot the diphoton invariant mass distribution of \kpiggn
for three different values $a_{V,0} = \, -0.4, \, -0.7, \, -1.0$. This
error interval is motivated either by the error in the slope of
\kggs or our uncertainty in $C_{-}(m_{\rho})$. Then correspondingly
to the values of $a_{V,0}$, 
\begin{equation}
Br( \kpiggne ) \, = \, \left\{ \begin{array}{cl}
                       1.12 \, \times \, 10^{-6} & , a_{V,0} = -0.4 \\
                       1.50 \, \times \, 10^{-6} & , a_{V,0} = -0.7 \\
                       2.06 \, \times \, 10^{-6} & , a_{V,0} = -1.0 
                               \end{array} \right. \; \; \; \; \; ,
\label{eq:brdiffer}
\end{equation}
which central value is in nice agreement with experiment. In Figs.~3 and
4 the small $m_{\gamma \gamma}$ region is interesting due to the 
dominance of the $B$ amplitude. Comparing
the curves for ${\cal O}(p^6)$  $a_{V,0}=0$ in Fig.~3 (dashed line)
with the one for \ops $a_{V,0}=-0.4$ in Fig.~4 (dotted line) we see that
there is a cancellation, for $a_{V,0}=-0.4$, between the
vector resonance exchange and unitarity contributions to the $B$ 
amplitude.

\begin{figure}
\begin{center}
\leavevmode
\hbox{%
\epsfxsize=14cm
\epsffile{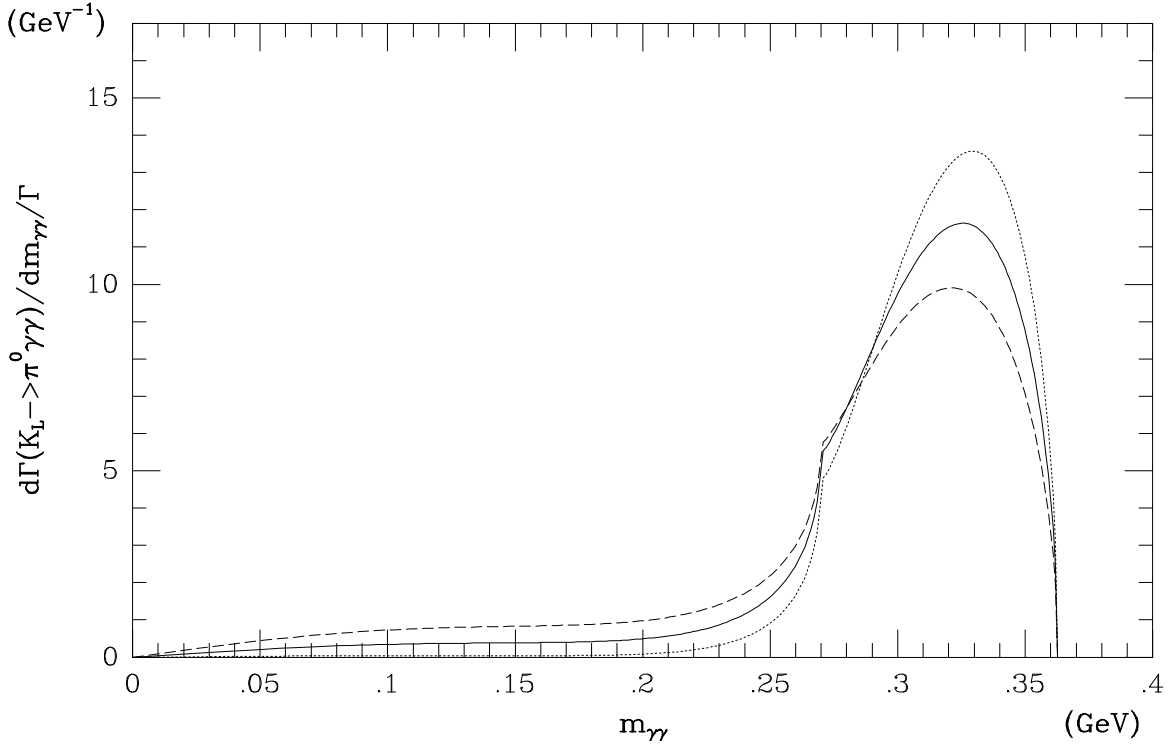}}
\end{center}
\vspace*{-10cm}
\caption{Normalized diphoton invariant mass spectrum for \kpiggn
at \ops (with the inclusion of the same contributions as in 
Fig.~3) with $a_{V,0} = -0.7$ (full line), $a_{V,0} = -0.4$ (dotted
line) and $a_{V,0} = -1.0$ (dashed line).}
\end{figure}

The discontinuity contribution of \kpiggn to the CP--conserving
amplitude of $K_L \rightarrow \pi^{0} e^+ e^-$ in the range of $a_{V,0}$
values considered above is
\begin{equation}
Br( K_L \rightarrow \pi^0 e^+ e^- ) \, |_{abs} \, = \, 
                       \left\{ \begin{array}{cl}
                       0.11 \, \times \, 10^{-12} & , a_{V,0} = -0.4 \\
                       1.24 \, \times \, 10^{-12} & , a_{V,0} = -0.7 \\
                       3.59 \, \times \, 10^{-12} & , a_{V,0} = -1.0 
                               \end{array} \right. \; \; \; \; \; .
\label{eq:eeabs}
\end{equation}
This result is consistent with the values obtained in \cite{CD93,DG95,SH92}.
\par
In \kpiggp the situation is rather different. First there is an unknown
counterterm parameter $\hat{c}$ Eq.~(\ref{eq:chat}) at \opc, and 
then all the models coincide in giving a small value for $a_{V,+}$ that
is not going to change the prediction we have already pointed out in 
\cite{DP96} where the unitarity correction of $K^+ \rightarrow \pi^+ 
\pi^+ \pi^-$ has been included.
\par
There is also another observable in \kpiggtot sensitive to the value
of $a_V$. This is the $y$--spectrum once the variable $z$ Eq.~(\ref{eq:yz})
has been cut in the upper limit. In Fig.~5 we show the normalized 
$y$--spectrum for \kpiggn and $z \, < \, 0.3$ for $a_{V,0} = 0,-0.4,-0.7, 
-1.0$. The dependence in $a_V$ is manifest. 
The interplay between \ops unitarity and vector meson exchange contributions
that, as mentioned above, may give destructive interference, is reproduced
also in Fig.~5. Indeed we notice that the behaviour of the curve for
$a_{V,0}=-0.4$ is essentially equal, though different in scale, to 
the one that arises at \opc \cite{EP90}. In Fig.~6 we show the normalized
$y$--spectrum for \kpiggp with $z \, < \, 0.3$ for $\hat{c}=0$ and 
$a_{V,+} = 0,-0.2$.

\begin{figure}
\begin{center}
\leavevmode
\hbox{%
\epsfxsize=14cm
\epsffile{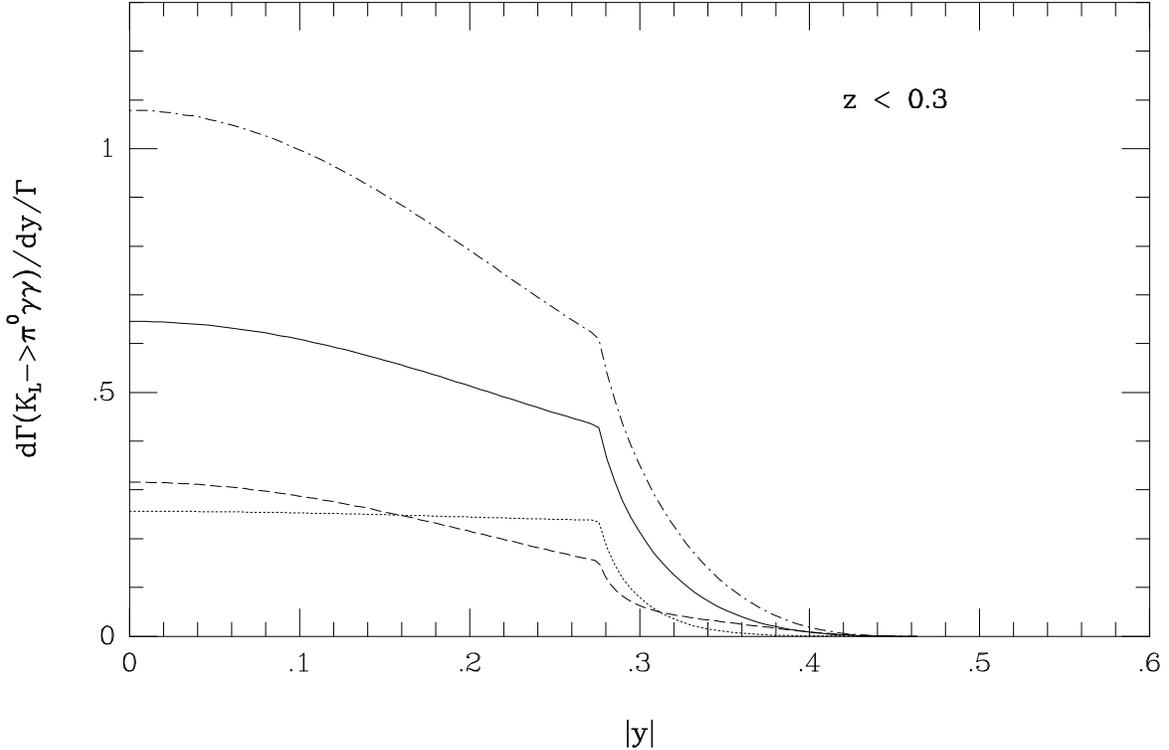}}
\end{center}
\vspace*{-9.5cm}
\caption{Normalized $y$--spectrum for \kpiggn and $z < 0.3$ for 
$a_{V,0}=0$ (dashed line), $a_{V,0}=-0.4$ (dotted line), 
$a_{V,0}=-0.7$ (full line), $a_{V,0}=-1.0$ (dash--dotted line).
No cut in $z$ is considered in the normalizing width. The corresponding
branching ratios are given in Eqs.~(\ref{eq:brklthe},\ref{eq:brdiffer}).}
\end{figure}

Ko has presented a rather involved model \cite{KO91} in order to evaluate
several vector meson dominated radiative kaon decays. It uses~: i) a 
realization of vectors in the Hidden symmetry formulation; ii) assumes
$\Delta I = 1/2$ enhancement, i.e. uses $G_8$ in Eq.~(\ref{eq:g8fr}),
 for the full lagrangian (including vectors);
iii) includes a free coupling in front of the second (and/or third) term
in Eq.~(\ref{eq:anomvpg}) that, according to the author, measures the
penguin contribution; iv) the numerically very important unitarity
corrections for \kpiggtot are not taken into account; v) uses nonet
symmetry in \kggs and \kpiggtot.
Ko concludes that the penguin contribution is noticeable,
result that is at odds with our conclusions and the idea of the BMS model~:
we have shown that absence of penguin contributions is 
compatible with no $\Delta I = 1/2$ enhancement in the processes 
considered. Any or several of the aforementioned points could be at the
origin of our different results.
Also only the octet scheme describes correctly according to us, at this order,
the phenomenology while the nonet does not. 

\begin{figure}
\begin{center}
\leavevmode
\hbox{%
\epsfxsize=14cm
\epsffile{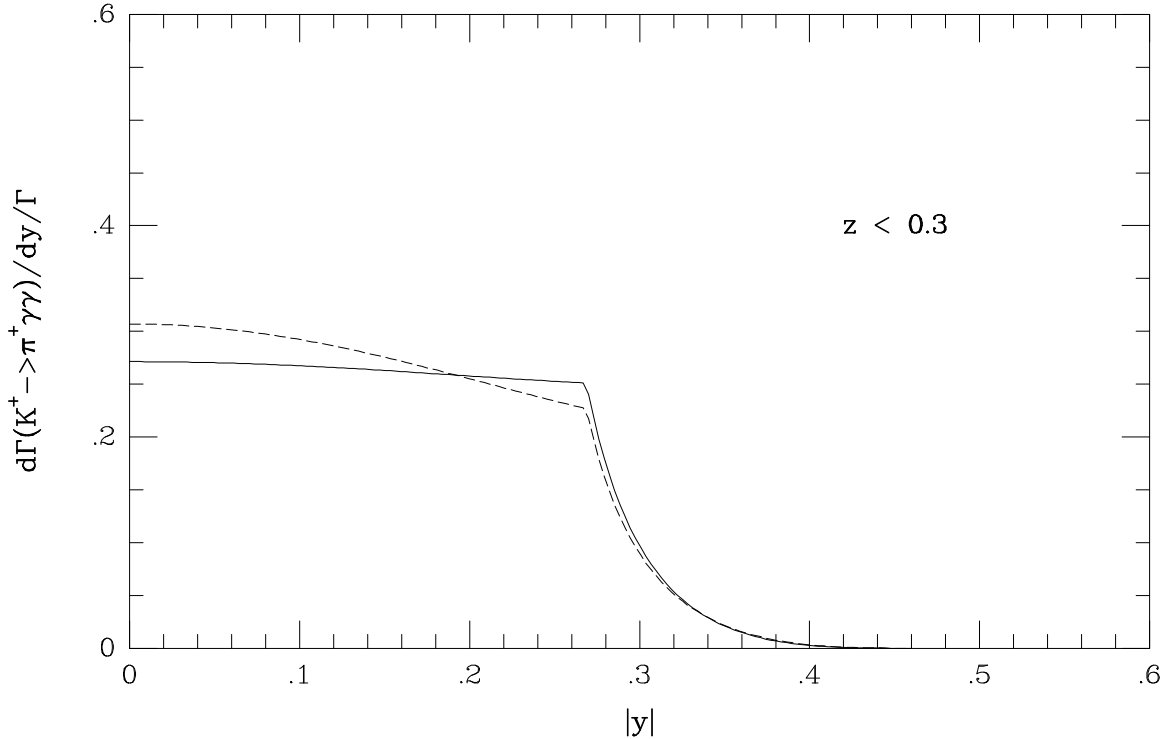}}
\end{center}
\vspace*{-10cm}
\caption{Normalized $y$--spectrum for \kpiggp with $\hat{c}=0$ and 
$z < 0.3$ for $a_{V,0}=0$ (dashed line) and $a_{V,0}=-0.20$ (full line).
The normalizing branching ratios (with no cut applied to $z$) are 
$7.24 \times 10^{-7}$ and $7.52 \times 10^{-7}$ respectively.}
\end{figure}

\begin{sidewaystable}
\begin{center}
\begin{tabular}{|c|c|c|c|c|c|c|c|c|c|}
\multicolumn{10}{c}{{\bf Table II}} \\
\multicolumn{10}{c}{} \\
\hline
\hline
\multicolumn{1}{|c|}{} &
\multicolumn{1}{|c|}{} &
\multicolumn{4}{|c|}{$b_D$} &
\multicolumn{3}{|c|}{$b \, = \, b_V \, + \, b_D$} &
\multicolumn{1}{|c|}{} \\ \cline{3-9} 
Currents & $b_V$ & & & & & & & & $b_{exp}$  \\
& & WDM & FM & BMS & FMV & WDM & FM & BMS $\,  + \,$ FMV & \cite{OA90} \\
\hline
\hline
& & & & & & & & & \\
octet (no   $\eta'$) & 0 & 0.35 & 0.71 $k_F$ & 0.3-0.4 & 0.51 &
0.35 & 0.71 $k_F$ & 0.8-0.9 & \\
& & & & & & & & & \\
\cline{1-9} & & & & & & & & & 0.81 \, $\pm$ \, 0.18 \\ 
& & & & & & & & & \\
nonet  ($\eta'$) & 0.46 & $--$ & 1.41 $k_F$ &  0.3-0.4 & 0.66 & 
$--$ & 0.46 $+$ 1.41 $k_F$  & 1.4-1.5 & \\
& & & & & & & & & \\
\hline
\hline 
\end{tabular}
\caption{Results for the slope $b$ of \kggs in the WDM, FM, BMS
and FMV models.
The result of the FMV model is for $\eta \, = \, 0.21$ as given in 
Eq.~(\ref{eq:eta8}). For the slope predicted by WDM in the nonet case
see the discussion in Subsection 4.4.}
\end{center}
\end{sidewaystable}

\subsection{\kggs}
\hspace*{0.5cm}
We have given two results for $b_D$ in our FMV model Eq.~(\ref{eq:bstaranom})
according to the consideration of including the $\eta'$ (nonet currents)
or not (octet currents). 
We have collected in Table II the results that we subsequently explain. 
\par
If we consider the octet case, $b^{octet}_V = 0$ 
and adding the result of BMS and FMV models we get 
\begin{equation}
b_{theo}^{octet} \, \simeq \, 0.8 - 0.9 \; \; \; ,
\label{eq:btheoro}
\end{equation}
to be compared with the experimental result Eq.~(\ref{eq:bexpsl}) 
$b_{exp} = 0.81 \pm 0.18$,
while if we consider the inclusion of nonet currents, $b_V$ is given
by Eq.~(\ref{eq:rvst}) and the final result is 
\begin{equation}
b_{theo}^{nonet} \, \simeq \, 1.4 - 1.5 \; \; \; ,
\label{eq:btheorn}
\end{equation}
that looks definitely too large. However the inclusion of the $\eta'$
amounts to the inclusion of a higher chiral order and 
other effects could be relevant.
\par
The correlation between the slope $b_D$ in \kggs and the 
$a^{dir}_{V,0}$ parameter in \kpiggn can be nicely quantified in the
FM (by eliminating $k_F$ between $b_D^{octet}$ in Eq.~(\ref{eq:avfm}) and
Eq.~(\ref{eq:bkstfm})) or in the FMV model (by carrying our results
for $\kappa_i$ couplings in Eq.~(\ref{eq:kappanom1}) into 
Eq.~(\ref{eq:comparaob})). The result happens
to be the same in both cases and we get
\begin{equation}
a^{dir}_{V,0} \, = \, - \, 4 \, \sqrt{2} \, \pi \,  
\Frac{h_V}{f_V} \, 
\Frac{|A_{\gamma \gamma}^{exp}|}{G_8 \, F_{\pi} \, \alpha_{em}}
\, b_D^{octet} \; \; \; \; .
\label{eq:comparison}
\end{equation}
We stress that $a_{V,0}^{dir}$ and $b_D^{octet}$ in this equation
only account for the FM or FMV model contributions. In our approach
the BMS model contribution should be added up to the FMV model.
\par
It is also  necessary to comment that, at \ops, there is a loop
contribution for $P \rightarrow \gamma \gamma^*$ that we are not including.
However this correction is known to be small \cite{BB85} and therefore
is not due to change our final result for $b$ more than a $10\%$.
\par
Bringing back the discussion about our coupling $G_8^{eff}$ that we 
consider the main source of model dependence, we have to stress that
a value of $G_8^{eff}$ much different than the one prescribed by the
Wilson coefficient at $\mu = m_{\rho}$ Eq.~(\ref{eq:eta8}) would
spoil the slope of \kggs and therefore, {\em a posteriori}, this
observable puts a constraint over $G_8^{eff}$.
\par
In any case we have already commented above that the experimental 
result for the $b$ slope Eq.~(\ref{eq:bexpsl}) is obtained through a fit
to the BMS model and doing an expansion in the $x$ variable. We consider
that this procedure might underestimate the slope.

\section{Conclusions}
\hspace{0.5cm} 
The \ops vector exchange contributions to the channels \kpiggn, 
\kpiggp and \kgll are tightly correlated due to the structure of the
weak $VP\gamma$ vertex. 
\par
In \kpiggtot while \kpiggp is still poorly known from the
experimental point of view (a situation due to change soon with the 
first events already collected by BNL-787 and the foreseen experiment
KLOE at DA$\Phi$NE), the situation of \kpiggn is much better.
As concluded in \cite{CE93} it seems that \kpiggn needs a noticeable local 
contribution
from vector exchange in order to bring agreement between
theory and experiment. 
\par
The slope of \kggs determined experimentally from \kgee can be computed
using octet or nonet symmetry ($\eta'$ on the same footing of $\eta$). 
We have pointed out that, for consistency, one should use the same
symmetry scheme in $b_D$ and $b_V$.
\par
Nonet symmetry in the BMS model gives a good result for the slope but
a too small $a_{V,0}$ and, consequently, does not accomplish our 
criterium of a simultaneous description of both \kggs and \kpiggn. The
WDM might describe the slope but probably gives a too small value for
$a_{V,0}$ and, in any case, the dynamical mechanism underlying the theory
is poorly known. The FM could describe well both the slope and 
$a_{V,0}$ but for a value $k_F \simeq 1$, that is, naive FM.
This would imply that the enhancement of the $\Delta I=1/2$ is at work 
in these channels. However our study
shows that this fact is a fake of the application of the FM because, as
we have found out, there is an
important factorizable contribution that was missing.
\par
We have first proposed a framework in which the full structure of the
most general octet weak $VP\gamma$ vertex (with the octet of pseudoscalars) is
presented. This allows us to parameterize the observables $a_V$ and $b_D$
in terms of five unknown effective coupling constants. Then we have applied
the FM in a novel approach: we use it in order to carry the construction
of the weak $VP\gamma$ vertex instead of the weak $K\pi\gamma\gamma$ or
$K\gamma\gamma^*$ vertices. In this way we discover a new chiral structure
contributing to both $a_V$ and $b_D$ that is missed in the usual approach
and without including any extra incertitude in the couplings. This 
contribution is due to the Wess--Zumino--Witten anomaly. This procedure
together with the choice of the effective coupling Eq.~(\ref{eq:eta8})
constitutes our Factorization Model in the Vector couplings (FMV). 
The effective
coupling Eq.~(\ref{eq:eta8}) amounts to the bare Wilson coefficient
in the non--leptonic hamiltonian and therefore no enhancement $\Delta I = 1/2$
is included in our model. As already pointed out the experimental
slope of \kggs might prefer a slightly larger value for the coupling 
but probably still compatible within the error of the Wilson coefficient.
\par
We emphasize also that it is very important to have a reliable and
predictive model to establish clearly other uncertainties involved in 
the study of these processes (higher order corrections, large
$SU(3)$ breaking \cite{HA96}, other resonance exchanges, etc.). 
Thus the effectiveness
of the Wilson coefficient in the description of the phenomenology of 
these processes, we think, is a relevant step forward for a predictive
description and understanding of non--leptonic kaon decays.
\par
In Tables I and II we collect our main results for $a_V$ and $b_D$ and 
we compare them with the rest of the analysed models. We conclude that
the FMV model with the new contribution that we have found gives a 
consistent picture of both \kpiggtot and \kggs processes and, in particular,
predicts a value of $a_{V,0} = -0.72$ in rather good agreement with the 
phenomenological estimate $a_{V,0} \simeq -0.8$. 
We have also considered the discontinuity contribution of \kpiggn to 
the CP--conserving amplitude of $K_L \rightarrow \pi^0 e^+ e^-$ for 
our value of $a_{V,0}$. The result, given in Eq.~(\ref{eq:eeabs}), agrees
with previous estimates.
Then we use our results in order to show the diphoton invariant mass
spectrum of \kpiggn including the \ops unitarity corrections from 
$K_L \rightarrow \pi^{0} \pi^+ \pi^-$, the experimental amplitude
$\gamma \gamma \rightarrow \pi^{0} \pi^{0}$ and the vector
meson contribution we have evaluated (Fig.~3 and Fig.~4). Also shown in 
Fig.~5 and Fig.~6 are the $z$--cut $y$--spectrum from \kpiggn and \kpiggp ,
respectively, that
happen to be sensitive to the $a_V$ parameter \cite{EP90} and therefore
worth to get from the experiments in order to clarify the subject.
\par
For completeness we have also shown in the Appendix that our model 
can also be derived in the Hidden Symmetry formulation of vector mesons
\cite{BK88} where the phenomenological couplings are recovered in the
so--called \lq\lq complete vector meson dominance scheme".
\vspace*{0.5cm} \\
{\large \bf Acknowledgements}
\vspace*{0.3cm} \\
\hspace*{0.4cm}
The authors thank G. Ecker, G. Isidori and A. Pich for very fruitful 
discussions. We also thank L. Cappiello for suggestions.
J.P. is supported by an INFN Postdoctoral fellowship.
J.P. is also partially supported by DGICYT under grant PB94--0080.
\vspace*{0.7cm} \\

\appendix
\newcounter{vector}
\renewcommand{\thesection}{\Alph{vector}}
\renewcommand{\theequation}{\Alph{vector}.\arabic{equation}}
\setcounter{vector}{1}
\setcounter{equation}{0}
\hspace*{-0.6cm}{\large \bf Appendix}
\vspace*{0.5cm} \\
\hspace*{0.5cm}
We have constructed the FMV model in the frame of $\chi$PT using the 
Callan--Coleman--Wess--Zumino formalism \cite{CCW69} for the pseudoscalar
sector and the inclusion of vector mesons as matter fields with a vector
field realization. We commented that the antisymmetric realization of 
the vector fields is less suitable in the odd--intrinsic parity sector
due to simple kinematical reasons \cite{EP90}. However we have found 
out that the
realization of vector mesons in the Hidden Symmetry model \cite{BK88}
is as well behaved as our formulation. Its application to the study
of the odd--intrinsic parity violating strong lagrangian was developed
in \cite{BB85,FK85}.
\par
The Hidden Symmetry model relies on the property that any nonlinear
sigma model based on the manifold $G/H$ is known to be gauge equivalent
to a \lq\lq linear" model with $G_{global} \otimes H_{local}$ symmetry, and
the gauge bosons corresponding to the hidden local symmetry, $H_{local}$
are composite fields. In \cite{BK88} was proposed that vector mesons
are to be identified with the dynamical gauge bosons of hidden local
$U(3)$ symmetry in the $U(3)_L \otimes U(3)_R \, / \, U(3)_V$ nonlinear
sigma model.
\par
The ideally mixed vector nonet $\rho_{\mu}$,
\begin{equation}
\rho_{\mu} \, = \, \Frac{1}{\sqrt{2}} \, \sum_{i=1}^8 \, \rho_{\mu}^i \, 
\lambda_i \, + \, \Frac{1}{\sqrt{3}} \, \rho_{\mu}^0 \; \; \; \; ,
\end{equation}
now transforms inhomogeneously under the chiral group as
\begin{equation}
\rho_{\mu} \, \mapright{G} \, h \rho_{\mu} h^{\dagger} \, + \, 
i \, h \partial_{\mu} h^{\dagger} \; \; \; , \; \; h \in U(3)_V
\end{equation}
in contradistinction with our realization in Eq.~(\ref{eq:vmhvh}). 
\par
The Lagrangian describing the vector mesons (to be added to 
${\cal L}_2$ in Eq.~(\ref{eq:str2})) is
\begin{eqnarray}
{\cal L}_{\rho} \,&  = & \, a \, F_{\pi}^2 \, \langle \, ( \, i \, 
\sigma^{\dagger} \, D_{\mu} \, \sigma \, )^2 \, \rangle \, - \, 
\Frac{1}{4} \, \langle \, \rho_{\mu \nu} \, \rho^{\mu \nu} \, \rangle
\; \; \; \; , \nonumber \\
i \, \sigma^{\dagger} \, D_{\mu} \, \sigma \, & \equiv  & \, 
i \, \sigma^{\dagger} \partial_{\mu} \sigma \, - \, 
g \, \sigma^{\dagger} \rho_{\mu} \sigma \, + \, v_{\mu} \; \; \; \; , 
\label{eq:lrhoh} \\
\rho_{\mu \nu} \, & \equiv  & \, \partial_{\mu} \rho_{\nu} \, - \, 
\partial_{\nu} \rho_{\mu} \, + \, i \, g \, [ \, \rho_{\mu} \, , \, 
\rho_{\nu} \, ] \; \; \; \; , \nonumber \\
v_{\mu} \, & \equiv  & \, \Frac{1}{2i} \, \left[ \, 
\xi^{\dagger} \, ( \, \partial_{\mu} \, - i \, r_{\mu} \, ) \, \xi \, 
+ \, \xi \, ( \, \partial_{\mu} \, - \, i \, l_{\mu} \, ) \,
\xi^{\dagger} \,  \right] \; \; \; \; , \nonumber  
\end{eqnarray}
where $a$ is a free parameter in the model and $\sigma$ is a 
$3 \times 3$ unitary matrix of unphysical scalar fields. These scalars 
give the longitudinal component to the 
$3 \times 3$ vector meson matrix $\rho_{\mu}$ and will be gauged away 
fixing the unitary gauge where $\sigma$ is the $3 \times 3$ identity
matrix. 
\par
The relevant Lagrangian for our $PV\gamma$ vertices, that will 
generate the analogous to $S(VP\gamma)$ in Eq.~(\ref{eq:actions})
turns out to be \cite{BB85}
\begin{eqnarray}
{\cal L}_{odd}^H \, & = & \, \Frac{i}{2} \, 
\varepsilon^{\mu \nu \alpha \beta} \, \left[ \, g \, a_2 \, 
\langle \, \sigma^{\dagger} \, \rho_{\mu \nu} \, \sigma \, 
\{ \, a_{\alpha} \, , \, \sigma^{\dagger} \, D_{\beta} \, \sigma \, \} \, 
\rangle \right. \nonumber \\
& &  \qquad \qquad  \, \left. - \, a_3 \, \langle 
\, ( \, \xi \, F_{\mu \nu}^L \, \xi^{\dagger} \, + \, 
\xi^{\dagger} \, F_{\mu \nu}^R \, \xi \, ) \, \{ \, a_{\alpha} \, , 
\, \sigma^{\dagger} \, D_{\beta} \, \sigma \, \} \, \rangle \, \right] \; \; .
\end{eqnarray}
In the unitary gauge $\sigma \rightarrow I$ and then $\xi \rightarrow u$
and $a_{\mu} \rightarrow u_{\mu}$ defined in Eq.~(\ref{eq:extrasym}).
Three new coupling constants appear in ${\cal L}_{odd}^H$~: $g$, 
$a_2$ and $a_3$. 
$g$ is the gauge coupling associated to the hidden symmetry and it can
be evaluated from $\Gamma(\rho \rightarrow \pi \pi)$ as $|g| \, = \, 
m_{V}/(2 F_{\pi}) \, \simeq \, 4.1$, $a_2$ and $a_3$ can be determined
from the strong $V \rightarrow P \gamma$ processes. The phenomenology
of these and Vector Meson Dominance is consistent with the
choices \cite{BB85}
\begin{eqnarray}
a_2 \, & = & \, 2 \, a_3 \, \simeq \, - \, \Frac{3}{16 \pi^2} \; \; \;  , 
\nonumber \\
a \, & = & 2 \; \; \; \; \; ,
\end{eqnarray}
(see however \cite{BGP95}) which induces the so--called 
\lq\lq complete vector meson dominance" \cite{FK85}.
From ${\cal L}_{odd}^H$ and comparing with 
Eq.~(\ref{eq:vpga}) we can express $h_V$ in terms of the couplings
appearing in this formalism. We get 
\begin{equation}
h_V \, = \, - \, \Frac{1}{2} \, g \, a_2 \; \simeq 3.9 \, \times 10^{-2} 
\; \; , 
\end{equation}
in very good agreement with the phenomenological value $|h_V| \, = \, 
(3.7 \pm 0.3) \times 10^{-2}$.
\par
In the Hidden Symmetry model the coupling $\rho \gamma$ is generated by the
first term of ${\cal L}_{\rho}$ in Eq.~(\ref{eq:lrhoh}).
By comparison and using the relation $m_V = 2 g F_{\pi}$ (that also
arises in the model) we find \cite{EGL89}
\begin{equation}
f_V \, = \, \Frac{1}{\sqrt{2} \, g} \, \simeq \, 0.17 \; \; \; \; , 
\end{equation}
to compare with the phenomenological value $|f_V| \simeq 0.20$. We note
that $h_V f_V > 0$ as indicated by phenomenology. See Section 2.1.
\par
From Eq.~(\ref{eq:rhov}), the expressions for $h_V$ and $f_V$ and the
value of $a_2$ we get
\begin{equation}
\ell_V \, = \, - \Frac{1}{\sqrt{2}} \, a_2 \, f_V \, 
                \Frac{m_V^2}{F_{\pi}^2} \, = \, 4 \, h_V \; \; \; \; \; , 
\end{equation}
verified phenomenologically within an error of $\sim 20 \%$.
\par
Therefore, analogously to Eq.~(\ref{eq:actions}) we can construct
the effective actions of our interest in this model and then 
carry out the same application of our FMV model. As commented in 
the text the results are consistent in both procedures.

\end{document}